\documentclass[preprint,aps,prd,amsmath,amssymb]{revtex4}
\usepackage{graphicx}
\usepackage{color}
\usepackage[latin1]{inputenc}
\newcommand{\be}{\begin{eqnarray}}
\newcommand{\ee}{\end{eqnarray}}
\newcommand{\nn}{~\nonumber \\}

\newcommand{\bmp}{\noindent\begin{minipage}{16cm}}
\newcommand{\emp}{\end{minipage}\vskip 7mm} 

\usepackage{graphicx}
\usepackage{dcolumn}
\usepackage{bm}
\usepackage{amsmath}
\usepackage{amsfonts}
\usepackage{bbm}
\usepackage{subfigure} 

\newcommand{\drawsquare}[2]{\hbox{%
\rule{#2pt}{#1pt}\hskip-#2pt
\rule{#1pt}{#2pt}\hskip-#1pt
\rule[#1pt]{#1pt}{#2pt}}\rule[#1pt]{#2pt}{#2pt}\hskip-#2pt
\rule{#2pt}{#1pt}}
\newcommand{\Yfund}{\raisebox{-.5pt}{\drawsquare{6.5}{0.4}}}
\newcommand{\Ysymm}{\Yfund\hskip-0.4pt%
                    \Yfund}
\def\symm{\Ysymm}
\def\bsymm{\overline{\Ysymm}}
\def\drawbox#1#2{\hrule height#2pt
        \hbox{\vrule width#2pt height#1pt \kern#1pt
              \vrule width#2pt}
              \hrule height#2pt}
\def\Fund#1#2{\vcenter{\vbox{\drawbox{#1}{#2}}}}
\def\Asym#1#2{\vcenter{\vbox{\drawbox{#1}{#2}
              \kern-#2pt 
              \drawbox{#1}{#2}}}}

\def\fund{\Fund{6.4}{0.3}}

\def\bfund{\overline{\fund}}

\begin{document}
\title{Light Composite Higgs from Higher Representations\\ versus \\Electroweak Precision Measurements\\ --- \\ Predictions for LHC}
\author{Dennis D. {\sc Dietrich}}
\email{dietrich@nbi.dk} 
\author{Francesco {\sc Sannino}}
 \email{francesco.sannino@nbi.dk}
\affiliation{The Niels Bohr Institute, Blegdamsvej 17, DK-2100 Copenhagen \O, Denmark }
 \author{Kimmo {\sc Tuominen}}\email{kimmo.tuominen@phys.jyu.fi}
\affiliation{Department of Physics,  P.O. Box 35,
FIN-40014 University of Jyv\"askyl\"a, Finland\\
Helsinki Institute of Physics, P.O. Box 64, FIN-00014 University
of Helsinki, Finland }


\begin{abstract}
We investigate theories in which the technifermions in higher dimensional representations of the technicolor gauge group 
dynamically break the electroweak symmetry of the standard 
model. Somewhat surprisingly, for the two-index symmetric representation of the gauge group
the lowest number of techniflavors needed to render the underlying gauge theory quasi conformal is two. This is 
exactly one doublet of technifermions with respect to the weak interactions promoting these theories to ideal candidates of ``walking'' type technicolor models. 
{}From the point of view of the 
weak interactions the two techniflavor theory has a Witten anomaly, which we cure by introducing a fourth family of leptons. An elegant feature of this model is 
that the techniquarks resemble an extra family of quarks arranged in 
the two-index symmetric representation of the SU(2) technicolor theory. We have also studied the theory with three 
technicolors and two techniflavors in the two-index symmetric representation
of the gauge group, which does not require a fourth family of leptons. We confront the models with the recent 
electroweak precision
measurements and demonstrate that the two technicolor theory is a valid candidate for electroweak symmetry breaking 
via new strong interactions.

We investigate different hypercharge assignments for the two color theory and associated fourth family type of leptons. We find 
that the two technicolor theory is  within the ninety percent confidence level contours defined by the 
oblique parameters. The electroweak precision measurements provide useful constraints on the relative mass splitting of the 
new leptons. In the case of a
fourth family of leptons with ordinary lepton hypercharge the new heavy neutrino can be a natural candidate of cold dark matter. The mass of this heavy neutrino, constrained by the precision measurements, is slightly larger than the one of the neutral electroweak gauge boson while the charged lepton has a mass roughly twice as large as  the associated neutrino. Extensions with a larger value of the hypercharge 
naturally featuring doubly charged leptons are also favored by precision measurements.

{ We also propose theories in which the critical number of flavors needed to enter the conformal window is higher than the one with 
 fermions in the two-index symmetric representation, but lower than in the walking technicolor theories with 
 fermions only in the fundamental representation of the gauge group. The simplest theories are the ones in which 
 we add a (techni)gluino to the theory, while the rest of the  matter fermions remain in the fundamental representation of the gauge group. 
Although these theories share some features with split supersymmetric theories, they are introduced here to address the hierarchy problem.}
 
 Due to the near conformal/chiral phase transition, we show that the composite Higgs is very light compared to the intrinsic scale 
of the technicolor theory. {}For the two technicolor theory we predict the composite Higgs mass not to exceed $150$~GeV. {We also provide estimates for the Higgs mass in walking technicolor theories with fermions in the fundamental representation and show that in this case the mass is around $400$~GeV.} 
\end{abstract}


\maketitle

\section{Introduction}
The precise nature of the Higgs boson is without any doubt one of the most important problems of theoretical physics. Is the Higgs boson, if it exists at all, elementary or composite? The Large Hadron Collider experiment at CERN will be soon shedding light on this sector of the electroweak theory. The outcome will have a profound impact on our knowledge of Nature.

Already now much indirect information on the Higgs boson can be extracted from precision measurements of the weak interactions.
The minimal standard model of particle interactions, defined as the $SU(3)\times SU(2)\times U(1)$ gauge theory of quarks and leptons with a 
single elementary Higgs field, fits reasonably well the bulk of precision data \cite{Eidelman:2004wy}. The data indicate a preference for a small Higgs mass. The central value of the global fit result, $M_H=113^{+56}_{-40}$GeV, is slightly below the direct lower bound $M_H\geq 114.4$GeV. Including the results of the direct searches drives the 95\% upper limit to $M_H\leq 241$GeV.

However, as explained in the review paper by Peskin and Wells \cite{Peskin:2001rw}, it is likely that in order to correctly explain the electroweak symmetry breaking in the standard model new physics is required. New physics may strongly affect the electroweak fit and weaken some of the constraints. However, any new model devised to describe the electroweak symmetry breaking in the standard model is subject to precision electroweak constraints linked to the vacuum polarizations of the electroweak gauge bosons, and encoded in the oblique parameters 
\cite{Kennedy:1988sn,Peskin:1990zt,Peskin:1991sw,
Kennedy:1990ib}, see also \cite{Altarelli:1990zd,Marciano:1990dp} for different parameterizations and \cite{Barbieri:2003pr,{Chen:2005jx}} for 
new developments. In practice, the overall corrections induced by the new physics may be comparable to the ones induced by a light elementary Higgs boson \cite{Holdom:1990tc,{Golden:1990ig}}.

Of particular interest to us are models of electroweak symmetry breaking via new strongly interacting theories of technicolor type \cite{TC}. This is a mature subject (for recent reviews see \cite{Hill:2002ap,{Lane:2002wv}}) where considerable 
effort has been made to construct viable models. One of the main difficulties in constructing such extension of the standard model is the very limited knowledge about generic strongly interacting theories. This has led theorists to consider specific models of technicolor which resemble ordinary quantum chromodynamics and for which the large body of experimental data at low energies can be directly exported to make predictions at high energies. According to Peskin and Wells \cite{Peskin:2001rw} generic theories of composite Higgs contain larger corrections with respect to the minimal standard model, similar to those of a heavy elementary Higgs boson \cite{Peskin:1991sw}. In order to construct models of this 
kind (i.e. with a heavy Higgs boson) compatible with precision data one needs to introduce new ingredients compensating for these corrections. However, a heavy composite Higgs is not always 
an outcome of strong dynamics \cite{{Sannino:2004qp},Hong:2004td}. Here we are not referring to models in which the Higgs is a quasi Goldstone boson \cite{Dimopoulos:1981xc} which have been 
investigated recently \cite{Arkani-Hamed:2001nc}.

{ Some of the problems of the simplest technicolor models, such as
providing ordinary fermions with a mass, are
alleviated when considering new gauge dynamics in which the
coupling does not run with the scale but rather walks, i.e.
evolves very slowly \cite{Holdom:1981rm,{Yamawaki:1985zg},{Appelquist:an}, MY}.
These strongly interacting theories with enough matter content to quasi stop the running of the coupling constant as a function of the energy scale  have been named walking technicolor theories.
Most of the investigations in the literature used matter in the fundamental representation of the gauge group. In this case one needs a very large number of matter fields, roughly of the order of $4N$ with $N$ the number of technicolors  to achieve the walking. A large number of techniflavors has many shortcomings, such 
as large contributions to the oblique parameters and a very large number of unwanted Goldstone bosons.

However it was shown recently \cite{Sannino:2004qp,{Hong:2004td}} that it is possible to consider matter 
in higher dimensional representations and achieve walking for a very small number of fields. Due to the nature of these theories one was also able to make a connection, at a large number of colors, with a particular sector of super Yang-Mills theories (SYM) \cite{Armoni:2004uu}.

We note that the idea of using higher dimensional
representations of the gauge group for technicolor like theories is not a new one. 
{}For walking technicolor it was proposed by Eichten and Lane \cite{Lane:1989ej}. Corrigan and Ramond in \cite{Corrigan:1979xf} constructed an explicit model of 
technicolor with fermions in the two-index antisymmetric representation \footnote{Interestingly they have also proposed a different type of large number of 
color expansion in which the fermions are not kept in the fundamental representation of the gauge group.}. We have shown in \cite{{Sannino:2004qp},Hong:2004td} that the Corrigan and Ramond type of theories with sufficiently large 
number of flavors to be able to achieve walking are disfavored by precision measurements \cite{Hong:2004td}. 
}

In this paper we analyze in detail the phenomenological consequences of the theories introduced in \cite{Sannino:2004qp} and \cite{Hong:2004td}. 
We will show that the two color technicolor theory with two flavors in the two-index symmetric representation of the gauge group requires the existence of a new fourth family of leptons to avoid the Witten anomaly while still being consistent with all of the gauge anomalies. We study the effects on the precision measurements and show that this theory is still a simple and viable model. We 
will consider different possibilities for the choice of the hypercharge allowed by the gauge anomaly constraint as well as different 
ways of giving masses to new neutral leptons.

 We then consider the three color theory with still two flavors in the two-index representation of the gauge group for which no new lepton family is needed, and study the oblique corrections in this case.
 
A simple way to understand why the present theories are still viable is that the resulting composite Higgs is lighter than the one typically expected for conventional composite Higgs theories. 
This was already argued in \cite{Hong:2004td} using a relation with SYM for the one flavor sector of the theories we considered. Here we go beyond this simple estimate and show that when the flavor dependence is taken into account the Higgs mass is even lighter than the one predicted in \cite{Hong:2004td}. In order to determine the flavor dependence of the Higgs mass in our theories we use the fact that there is a chiral phase transition as function of the number of flavors, as well as trace anomaly arguments. Remarkably, the theories investigated here not only explain the hierarchy problem but also lead to a light composite Higgs.   
 
 { We also propose theories in which the critical number of flavors needed to enter the conformal window is higher than the one with 
 fermions in the two-index symmetric representation, but lower than the traditional walking technicolor theories with fermions 
 in the fundamental representation of the gauge group. A simple class of these theories are split (super)technicolor theories in which 
 we add only a (techni)gluino to the theory with still $N_{Tf}$ fermions in the fundamental representation of the gauge group. These 
 theories are closer to their supersymmetric parents \cite{Intriligator:1995au} and are similar to the ones used in 
 recent extensions of the standard model \cite{Arkani-Hamed:2004fb,{Giudice:2004tc},{Arkani-Hamed:2004yi}}. However, they are introduced 
 to address the hierarchy problem. {}For the phenomenology we focus on the technicolor theories with fermions in the higher representation of the gauge group.}
 
 The paper is structured as follows: In Section \ref{2} we briefly summarize the general features of the theories with fermions in 
the two-index symmetric and antisymmetric representation of a $SU(N)$ gauge group.  In Section \ref{3} we explore 
the two technicolor theory with two techniflavors. The Witten anomaly makes the introduction of a new lepton family necessary. We explore different hypercharge assignments as well as different ways of providing a mass to the neutral leptons. In Section \ref{4} we briefly introduce the three technicolor theory and summarize the generic properties of the technibaryon spectrum. 

In Section \ref{5} we confront our theories with electroweak precision measurements and derive constraints 
on the fourth generation of leptons. We show here that the two color technicolor theories with fermions 
in the two-index symmetric representation of the gauge group are viable candidates for breaking the electroweak symmetry dynamically. We also explore the constraints for the three technicolor theory. We 
analyze the properties of the fourth family of leptons with respect to the precision measurements in Section \ref{6}. 

In Section \ref{7} we demonstrate that the composite Higgs is naturally light. This is due to two relevant factors: 
1) The fermions are in the two-index symmetric representation of the gauge group. This fact allows us to relate the one (techni)flavor sector of the theory with SYM at large N and to demonstrate that the fermion-antifermion scalar in this theory is lighter than in QCD-like theories \cite{Hong:2004td}; 2) Using the trace anomaly and the fact that we are near a quantum phase transition as a function of the number of (techni)flavors, we determine the flavor dependence of the mass of the scalar fermion-antifermion. Normalizing our results for the mass of this scalar to the one flavor case we extrapolate to the 
two flavor case. Due to the nearby phase transition (as well as conformal fixed point), the mass of this scalar is 
reduced further with respect to the one flavor case. 
For the two techniflavor case we then predict the mass of the Higgs (i.e. the scalar) to be lighter than $150$~GeV.

Since the theories presented here are non-supersymmetric our non-perturbative predictions can be tested via current lattice simulations. It would be very interesting also to compare the lattice results for the spectrum of these theories. 
The technihadronic spectrum is important since it can be, in principle, observed 
at LHC.   

We then conclude in Section \ref{8} by summarizing the basic results and predictions while outlining future directions.

\section{Features of Higher Representations} 
\label{2}

\subsection{The Phase Diagram}
The simplest technicolor model has $N_{T f}$ Dirac fermions in
the fundamental representation of $SU(N)$. These models, when
extended to accommodate the fermion masses through the extended technicolor interactions,
suffer from large flavor changing neutral currents. This problem is
alleviated, at least to the extent of accounting for masses up to
that of the b quark, if the number of flavors is sufficiently large
such that the theory is almost conformal. This is estimated to happen
for $N_{T f} \sim 4 N$ \cite{Yamawaki:1985zg}, which implies a large
contribution to the oblique parameter $S$ (at least in perturbative
estimates) \cite{Hong:2004td}. Near the conformal window \cite{Sundrum:1991rf,
{Appelquist:1998xf}} the $S$ parameter is reduced due to
non-perturbative corrections, but might still be too large if the
model has a large particle content. In addition, such models may
have a large number of unwanted pseudo Nambu-Goldstone bosons. By
choosing a higher dimensional representation for the fermions one
can overcome these problems \cite{Sannino:2004qp,{Hong:2004td}}. {}For the reader's convenience 
we report below the basic features of the theories with two-index representations 
explored in \cite{Sannino:2004qp}.

These theories have
fermions in the two-index symmetric (S-type) or antisymmetric
(A-type) representation. In Table \ref{symmetric} we present the
generic S-type theory.
\begin{table}[h]
\begin{center}
\begin{minipage}{3in}
\begin{tabular}{c||ccccc }
 & $SU(N)$ & $SU_L(N_{T f})$& $SU_R(N_{T f})$&$U_V(1)$ & $U_A(1)$  \\
  \hline \hline \\
${T_L}$& $\symm$ & $\fund$ & $1$ & $1$ & $1$  \\
 &&&\\
 $\bar{T}_R$ & $\bsymm $ & $1$& $\bfund$& $-1$ & $1$ \\
 &&&\\
$G_{\mu}$ & {\rm Adj} & $0$&$0$ &$0$ & $0$    \\
\end{tabular}
\end{minipage}
\end{center}
\caption{Schematic representation of a generic nonsupersymmetric vector like $SU(N)$ gauge theory with matter content 
in the two-index representation.  Here $T_{L(R)}$ are Weyl fermions.}
\label{symmetric}
\end{table}
At infinite $N$ and with one Dirac flavor, the S- and A-type
theories become non-pertubatively equivalent to SYM
\cite{Armoni:2004uu}. This property was used to make predictions
for QCD with one flavor \cite{{Armoni:2004uu},Sannino:2003xe} or for the spectrum 
of SYM \cite{Feo:2004mr}. Using the results in \cite{Sannino:2003xe} one has been able in \cite{Hong:2004td} to argue that the composite Higgs in
these theories is expected to be lighter than in the typical theories used till now to break the electroweak symmetry dynamically. 
However these predictions were valid technically only for the one flavor technicolor theory of S-type. In this paper we will include the flavor corrections and show that when increasing the number of flavors the Higgs become even lighter than the one predicted in \cite{Hong:2004td}. 

The relevant feature, found in \cite{Sannino:2004qp}, is that the
S-type theories can be near conformal already at $N_{T f}=2$ when $N=2$ or $3$. This
should be contrasted with theories in which the fermions are in the fundamental
representation for which the minimum number of flavors required to
reach the conformal window is eight for $N=2$.

The $N=3$ model with A-type fermions is just $N_{T f}$-flavor QCD with the maximum allowed number of flavors equal to $16$. For $N=2$
the antisymmetric representation goes over to a pure Yang-Mills theory
with a singlet fermion. For S-type models, asymptotic freedom is
lost already for three flavors when $N=2$ or $3$, while the upper
bound of $N_{T f}=5$ is reached for $N=20$ and does not change when
$N$ is increased further.

\begin{figure}[t]
\includegraphics[width=16truecm,height=5truecm]{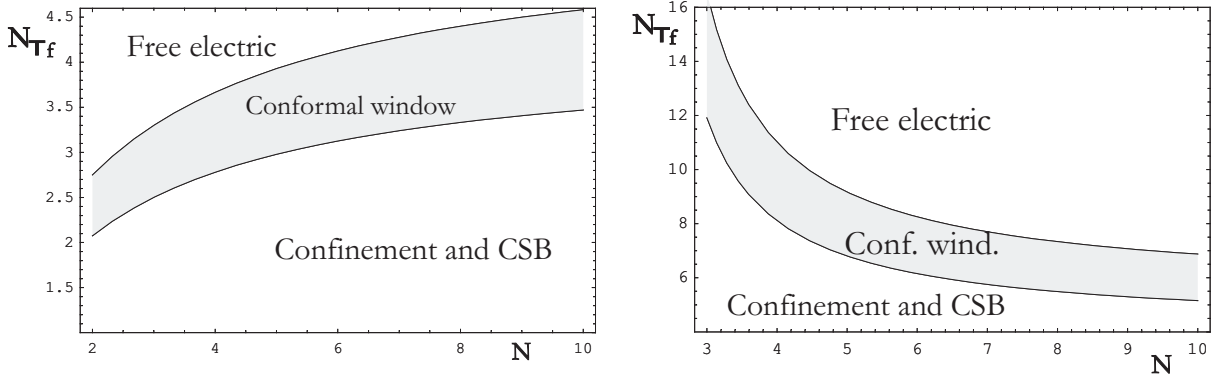}
\caption{Left(Right) panel: Phase diagram as function of number of $N_{Tf}$ 
Dirac
flavors and $N$ colors for fermions in the two-index symmetric (antisymmetric)
representation, i.e. S(A)-types, of the gauge group.} \label{PhaseDiagram}
\end{figure}

The phase diagram, studied in \cite{Sannino:2004qp} as a function of the number of colors and flavors
for the S- and A-type theories is summarized in figure \ref{PhaseDiagram}.
{}From the picture it is clear that for $N=2,3,4,5$ $N_{Tf}=2$ is already the highest
number of flavors possible before entering the conformal window.
Hence, for these theories we expect a slowly evolving coupling
constant. The critical number of flavors must be greater than three for $N\geq 6$, but
remains smaller than or equal to four for any $N$.

The critical value of flavors increases with the number of colors
for the gauge theory with S-type matter: the limiting value is
$4.15$ at large $N$. 
These estimates are based on the validity of the first
few terms in the perturbative expansion of the $\beta$-function. However, we will shortly review 
arguments provided in \cite{Sannino:2004qp}, which lead to the same prediction of the conformal window.

The situation is different for the theory with A-type matter. As is evident
from the phase diagram,
the critical number of flavors increases when decreasing the
number of colors. The maximum value of $N_{Tf}=12$ is obtained for
$N=3$, i.e. standard QCD. In reference \cite{Hong:2004td} it has been shown that 
the nearly conformal A-type theories have, already 
at the perturbative level, a very large $S$ parameter with respect to the experimental data. Hence we will 
not consider the A-type theories any further.

A better understanding of the non-perturbative dynamics of these
theories can be obtained by exploiting their relation with
supersymmetric Yang--Mills theory (SYM) \cite{Sannino:2004qp}.

As already mentioned, the conformal window of the S-theory is more robust than the corresponding one for a
generic non-supersymmetric theory. This is so for two reasons: 1) The
S-theory looses asymptotic freedom quite soon; already for $N_{Tf}=3$
or $N_{Tf}=4$ when $1<N\leq 5$ and for $N_{Tf} =5$ at large $N$. 2) The
$N_{Tf}=1$ sector of the theory (the orientifold field theory) at large
N is mapped into SYM which is known to confine. Hence we can exclude
the $N_{Tf}=1$ theory from being nearly conformal at large $N$. 

Clearly for $N > 3$ our prediction
is even more reliable. So we conclude that for $N=3$ the theory is
conformal or quasi-conformal already for $N_{Tf}=2$. We expect this result to
hold also for the case $N=2$. These arguments about the
conformal window are {\it completely} independent of the
non-perturbative methods used above while strengthening our results.

{Among other approaches the instanton-liquid model has also been used to investigate the QCD chiral 
phase transition as function of the number of flavors \cite{Schafer:1995pz}. Within that framework the critical 
number of flavors for chiral symmetry restoration was found to be substantially lower than $4N$ 
\cite{Schafer:1995pz}. In the future, it would be relevant to compare the predictions for the phase diagram 
(as function of the numbers of flavors and colors) of 
theories utilizing the two-index symmetric representation of the gauge group 
\cite{Sannino:2004qp}, with the ones for the instanton-liquid model \cite{Schafer:1995pz}. The relevant work in 
\cite{Grunberg:2000ap,{Gardi:1998ch},{Grunberg:1996hu}} can also be extended to the present theories.
Lattice simulations similar to the ones presented in \cite{Iwasaki:1996fk} would be interesting to compare with. }

The relation with SYM can also be employed to deduce some information about the
hadronic spectrum of non-supersymmetric
theories and vice versa \cite{Sannino:2003xe,{Feo:2004mr}}. This is possible since the $N_{Tf}=1$ bosonic sector of
the A/S-type theories, at large N, is mapped into the bosonic spectrum of SYM. In
\cite{Sannino:2003xe} and subsequently in \cite{Feo:2004mr} a number of relevant relations at
finite $N$ have been uncovered \footnote{
The knowledge of the scalar sector of
these theories should be confronted with the elusive one in ordinary
QCD \cite{Sannino:1995ik,Harada:2003em}, and
consequently in walking as well as in ordinary technicolor theories.}.

To generate fermion masss in technicolor models one needs
additional interactions arising from extended technicolor (ETC),
which couple technifermions to  ordinary
fermions~\cite{Lane:1989ej}. However, the ETC interaction
typically leads to unacceptably large flavor-changing neutral
currents. This problem is less severe if the technifermion
bilinear, whose condensate breaks electroweak symmetry, has a
large anomalous dimension as it happens in near conformal theories \cite{Holdom:1981rm,{Yamawaki:1985zg},{Appelquist:an}, MY}.

The enhancement of the condensate allows
reasonable masses for light quarks and leptons, even for
large ETC scales necessary to suppress
flavor-changing neutral currents sufficiently well. However, to obtain the observed top mass,
we must rely on additional dynamics, as in so-called non-commuting
ETC models, where the ETC interaction does not commute with the
electroweak interaction~\cite{Chivukula:1994mn}.%

If our goal is only to obtain an effective theory valid up to the
scale $\Lambda_{\rm ETC} \sim 10^3$ TeV, we need not explain the
origin of ETC operators (this is in the spirit of so-called
``little-Higgs'' models \cite{Arkani-Hamed:2001nc}, in which the Higgs is a quasi Goldstone boson \cite{Dimopoulos:1981xc}). We leave an explanation of quark
and lepton masses for future work.

Summarizing, the generic properties of the S-type theory are:
\begin{itemize}
\item[$\bullet$] {The conformal window starts at a low number of (techni)flavors. This number will never exceed four. For a small number of (techni)colors this number is slightly above 
two.}
\item[$\bullet$]{Near the conformal window two scales are generated. The one associated to the chiral 
symmetry breaking ($\Lambda_{TC}$) and the one above which the coupling constant starts running again ($\Lambda_{ETC}$). 
In the two and three (techni)color theory with two (techni)flavors:
\begin{eqnarray}
\Lambda_{ETC} > 300~\Lambda_{TC} \ .
\end{eqnarray}
Identifying $\Lambda_{ETC}$ with the natural scale for introducing the new dynamics generating the masses 
of the ordinary fermions one at the same time enhances the masses of the fermions and reduces the contribution 
from flavor changing neutral currents.}
\item[$\bullet$]{The one flavor sector of the theory at large number of (techni)color is mapped into SYM.}
\end{itemize}

\begin{figure}[t]
\includegraphics[width=10truecm,height=6truecm]{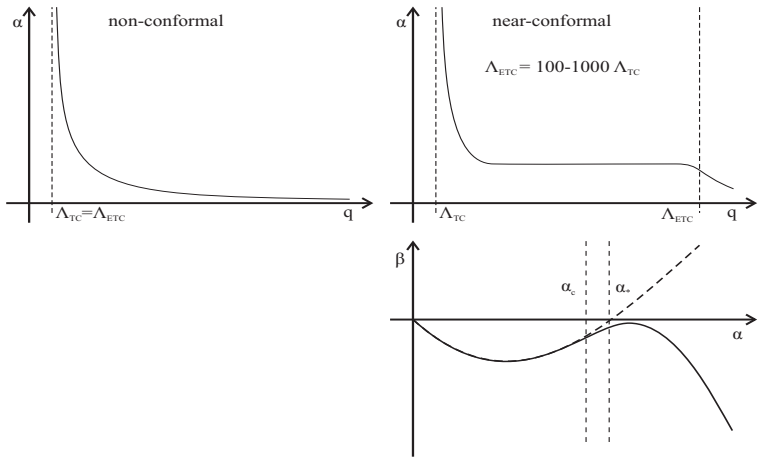}
\caption{Left Panel: A standard running behavior of a coupling constant in a generic asymptotically free theory. 
Right Panel: The walking behavior of the coupling constant when the number of flavors is near a conformal fixed point. 
The associated beta function is plotted below the previous graph. } \label{walking}
\end{figure}

\subsection{Possible alternatives: Split Technicolor}
If we insist on keeping the technifermions in the fundamental representation while trying to reduce the number of 
techniflavors needed to be near a conformal window one possibility is to add matter uncharged under the weak interactions. In this 
way one would, in general, increase the number of pseudogolstone bosons. However an interesting and minimal possibility is to consider 
adding a massless Weyl fermion in the adjoint representation of the gauge group. This particle behaves as a technigluino. The 
resulting theory has the same matter content as $N_{Tf}$-flavor super QCD but without the scalars. This split technicolor theory 
has the critical number of flavors above which one enters the conformal window lying within the range:
\begin{eqnarray} 
\frac{3}{2}< \frac{N^{c}_{Tf}}{N} \lesssim 4 \ .
\end{eqnarray}
Here $N$ is the number of colors. The lower bound is the exact supersymmetric point for a non-perturbative 
conformal fixed point \cite{Intriligator:1995au} while the upper 
bound is the one expected in the theory without a technigluino. The critical value of techniflavors is expected 
in the range of values presented above since super QCD has a larger matter content than split super QCD while 
QCD has a even lower matter content than 
split super QCD. Interestingly, this shows that with two colors the number of (techni)flavors needed to be near 
the conformal window in the split case is at least three, while for three colors more than five flavors are required. 
These values are still larger than the ones presented 
for theories with fermions in the two-index symmetric representation, although still lower than the ones used in 
walking technicolor theories with fermions only in the fundamental representation of the gauge group. It is 
useful to remind the reader that in supersymmetric theories the critical number of flavors needed to enter the 
conformal window does not coincide with the critical number of flavors required to restore chiral symmetry. The 
scalars in supersymmetric theories play an important role from this point of view. 
We note that a split technicolor-like theory has been used recently in \cite{Hsu:2004mf}, to investigate the 
strong CP problem. 

Split technicolor is a theory sharing some features with theories of split supersymmetry recently advocated and 
studied in \cite{Arkani-Hamed:2004fb,{Giudice:2004tc},{Arkani-Hamed:2004yi}} as possible extensions of the 
standard model. Clearly we have introduced split technicolor, differently from split supersymmetry, to address 
the hierarchy problem. We then do not expect new scalars to appear at energy scales higher than the one of the 
electroweak theory.   

In the present work we focus on theories with 
fermions in higher representation of the gauge group.
\section{Two Technicolors and two flavors:\vskip .1cm  Witten's Anomaly and a New Lepton Family}
\label{3}

Here, the gauge dynamics driving electroweak symmetry breaking consist
of two Dirac fermions in the two-index symmetric representation
of the $SU(2)$ gauge theory. 
This model is the one with the smallest
perturbative $S$-parameter \cite{Hong:2004td} which preserves the following
relevant feature: It is (quasi)conformal with {\it just} one doublet of
techinfermions \cite{Sannino:2004qp}. This naturally leads to a two
scale theory \cite{Cohen:1988sq}: The lowest scale is the one at which the
coupling constant becomes strong (i.e. the electroweak scale). The
other scale is defined to be the one above which the $SU(2)$ of technicolor gauge
coupling constant starts running. These two scales are exponentially
separated. The typical behavior of the coupling in these theories is schematically represented in figure \ref{walking}.
This fact allows us to concentrate on the physics near
the electroweak symmetry breaking scale. Using a bottom-up type of approach we postpone questions associated to the detailed
dynamics of the generation of fermion masses.

We represent the doublet of techni-fermions as:
\be
T_L^{\{C_1,C_2 \}}
=
\left(\begin{array}{l}U^{\{C_1,C_2 \}}\\ D^{\{C_1,C_2 \}}\end{array}\right)_L \ ,
\qquad 
T_R^{\{C_1,C_2\}}&=&\left(U_R^{\{C_1,C_2\}},~ D_R^{\{C_1,C_2\}}\right) \ .
\nn
\ee
Here $C_i=1,2$ is the technicolor index and $T_{L(R)}$ is a doublet (singlet) with respect 
to the weak interactions. 
The two-index symmetric representation of $SU(2)$ is real, and hence the global classical
symmetry group is $SU(4)$ which breaks to $O(4)$. This leads to the appearance of nine
Goldstone bosons, of which three become the longitudinal components of the weak gauge bosons.
The low energy spectrum is expected
to contain six quasi Goldstone bosons which receive mass through
extended technicolor (ETC) interactions \cite{Hill:2002ap,{Lane:2002wv},Appelquist:2002me,Appelquist:2003uu,{Appelquist:2003hn},{Appelquist:2004mn},{Appelquist:2004es},{Appelquist:2004ai}}.

As pointed out in \cite{Sannino:2004qp}, the weak interactions are also affected by the $SU(2)$ Witten anomaly
\cite{Witten:fp}. More specifically, since our techniquarks are in the two-index symmetric 
representation of $SU(2)$ we have exactly three extra left doublets from the point of view of the weak interactions. There can be different 
resolutions of this problem and all of them lead to new and interesting physical consequences observable at LHC. 
A simple way to cure such an anomaly without introducing further unwanted
gauge anomalies is to introduce at least one new lepton family. According to the choice of the hypercharge we discuss 
a number of relevant cases:

\subsection{New Standard Model Like Lepton Family}
\begin{center}
\begin{figure}[h]
\resizebox{!}{6.5cm}{\includegraphics{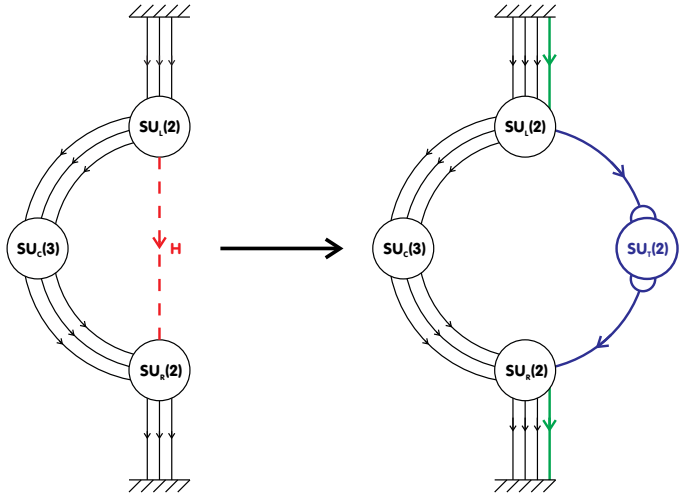}}

\caption{Schematic diagram summarizing the particle content 
and gauge interactions of the new minimal standard model like technicolor theory. On the left we represent the standard model. The oriented lines linking the walls to the 
$SU(2)_{L/R}$ circles are the leptons. The lines connecting the $SU(2)_{L/R}$ circles to $SU(3)_c$ are 
quarks. The dashed line is the ordinary Higgs field. On the right we have the same schematic picture except that now the Higgs is replaced by new gauge dynamics. Note that the oriented arrows terminate in the new 
$SU(2)$ of technicolor with two legs representing the two-index symmetric representation of SU(2). We 
have also added another left and right leptonic line. The new leptons are needed to avoid the global $SU(2)_L$ Witten anomaly. }
\label{schema}
\end{figure}
\end{center}

Since we have three doublets of techniquarks which resemble very much an ordinary triplet of colored quarks one can assign to the 
techniquarks 
the standard quark hypercharge which for the left-handed technifermions is then:
\begin{eqnarray}
Y=+1/6 \ .
\end{eqnarray}
The hypercharge is linked to the ordinary charge following the convention:
\begin{eqnarray}
Q = T_3 + {Y} \ .
\end{eqnarray}
 This yields:\be
T_L^{( Q)}
=
\left(\begin{array}{l} U^{(+2/3)}\\D^{(-1/3)}\end{array}\right)_L
\ee
where we have provided the electric charges of the techniquarks and suppressed the technicolor indices. {}For the right-handed techni-fermions which are isospin singlets we have:
\be
T_R^{(Q)}&=&\left( U_R^{(+2/3)},~ D_R^{(-1/3)}\right)
\nn
Y&=&~~+\frac{2}{3},~~~~-\frac{1}{3} \ .
\ee
In this case it is sufficient to add one new generation of left-handed leptons with hypercharge $Y=-1/2$:
\be
{\cal L}_L ^{(Q)}=
\left(\begin{array}{l}\nu_{\zeta}^{(0)} \\ \zeta^{(-1)}\end{array}\right)_L \ .
\ee
Clearly this new lepton family must be sufficiently heavy and not at odds with the electroweak precision measurements. 
We will consider three possible 
scenarios: The one in which the neutral and charged lepton have a Dirac mass. In the second we have a Majorana mass for the neutrino and a Dirac for the lepton. In the last scenario we provide a Dirac as well as a Majorana mass for the neutrino and of course always a Dirac mass for the charged lepton \cite{Bertolini:1990ek}. We will also barr any mixing between this fourth family of leptons and the three light families, however, it will be interesting in the future to investigate a more general mixing matrix with the light generations.
It is instructive to summarize the theory diagrammatically in figure \ref{schema}.
\subsubsection{Dirac Mass and a potential Dark-Matter candidate}
In this case we add one generation of right-handed leptons (isospin singlets)
\be
{\cal L}_R^{(Q)}&=&\left({\nu_{\zeta}}_R^{(0)},~\zeta_R^{(-1)}\right)
\nn
Y&=&~~~~0,~~-1 \ .
\ee
A standard Dirac mass term can be provided for the two leptons. If the neutral lepton is lighter than the associated charged lepton and since, by assumption, it does not mix with the lighter lepton generations, it becomes absolutely stable and a potential natural candidate for cold dark matter. This is so since a fourth stable neutrino has only weak interactions. Besides, we note that since the fourth family of leptons in our scenario is needed to compensate for the anomalies introduced by the techniquarks, its mass scale is naturally linked to the electroweak symmetry breaking scale. There are, however, a few caveats which need to be resolved to make the present neutrino a reasonable candidate for cold dark matter. Indeed if it turns out to be absolutely stable it will have to cluster \footnote{We thank D.J. Schwarz for 
suggesting this possibility to us.} \cite{Ringwald:2004np} rather than be distributed homogeneously in the universe as is assumed when deriving bounds on its mass \cite{reusser91,{Abusaidi:2000wg}}. In any case this neutrino can be 
a relevant component of cold dark-matter. Precision electroweak measurements, as we shall see, are already able to provide relevant information on the neutral versus charged lepton mass. 
\subsubsection{Majorana Mass}
{}From a theoretical point of view one can also provide a Majorana type mass for the new neutrino. In this case we need not to introduce the right neutrino field and the fourth family continuous lepton number is not conserved. The residual possible $Z_2$ symmetry under which only the neutrino transforms as
\begin{eqnarray}
{\nu_{\zeta}}_L \rightarrow -{\nu_{\zeta}}_L \ , \qquad \zeta_L \rightarrow \zeta_L \ ,
\end{eqnarray}
although left unbroken by the mass term, is violated by the weak interactions. {}For this type of heavy Majorana neutrino the bounds derived in \cite{reusser91} are much weaker even when assuming this type of matter does not cluster. It might then be a better candidate for cold dark matter.   
\subsubsection{Dirac \& Majorana Mass.}
This is the most democratic of all cases since we assign both a Dirac and Majorana mass to the neutrino. Following the work of Bertolini and Sirlin, we introduce a mass term in the Lagrangian of the type: \cite{Bertolini:1990ek}
\begin{eqnarray}
-m_{D} \left({\rho}^{\alpha}\xi_{\alpha} + \bar{\rho}_{\dot{\alpha}}\bar{\xi}^{\dot{\alpha}} \right) - \frac{M}{2} \left( \rho^{\alpha}\rho_{\alpha}+ \bar{\rho}_{\dot{\alpha}}\bar{\rho}^{\dot{\alpha}} \right) \ ,
\end{eqnarray}
with Weyl spinors
\begin{eqnarray}
\rho = {\nu_{\zeta}}_L \ ,\qquad  \bar{\xi} = {\nu_\zeta}_R \ , 
\end{eqnarray}
$\alpha=1,2$ are the spin indices and the dotted indices mean complex conjugated spinors \footnote{For the spinor 
notation we are following the Wess and Bagger notation \cite{WB}.}. One then diagonalizes the mass matrix obtaining two independent Majorana eigenstates with masses $M_1$ and $M_2$ related to the original Dirac and Majorana mass according to:
\begin{eqnarray}
M_1 M_2 = m^2_D \ , \qquad M_2-M_1 = M \ .
\end{eqnarray} 
 The original state can be rewritten in terms of the two Majorana eigenstates $N_1$ and $N_2$ as follows:
\begin{eqnarray}
\rho &=&  i \cos\theta P_L N_1 + \sin\theta P_L N_2 \nonumber \\
\xi &=& -i \sin\theta P_L N_1 + \cos\theta P_L N_2 \ . 
\end{eqnarray}
with $P_L=(1-\gamma_5)/2$ and 
\begin{eqnarray}
\tan2\theta = \frac{2m_D}{M} \ .
\end{eqnarray}
After Bertolini and Sirlin \cite{Bertolini:1990ek} had investigated the corrections to the custodial symmetry due to the presence of such a mass term for neutrinos Gates and Terning \cite{Gates:1991uu} computed the corrections for other relevant oblique parameters and showed that their 
contribution to the $S$ parameter can be negative. Unfortunately, with a single new lepton family, the absolute value is very small \cite{Gates:1991uu}.
\subsection{Fractionally Charged Leptons}
A solution of the anomalies is also obtained with the choice of the hypercharge $Y=0$ 
for the left handed technifermions:
\be
T_L^{(Q)}
=
\left(\begin{array}{l} U^{(+1/2)}\\ D^{(-1/2)}\end{array}\right)_L \ .
\ee
Consistency requires for the right-handed technifermions (isospin singlets):
\be
T_R^{(Q)}&=&\left(U_R^{(+1/2)},~D_R^{-1/2}\right)
\nn
Y&=&~~+1/2,~-1/2 \ .
\ee 
To avoid the Witten anomaly simultaneously while assuming only one new lepton family we must have for the left-handed leptons $Y=0$ as well
\be
{\cal L}_L^{(Q)}
=
\left(\begin{array}{l}\zeta^{(+1/2)} \\ \eta^{(-1/2)}\end{array}\right)_L \ ,
\ee
and for the right-handed leptons (isospin singlets):
\be
{\cal L}_R^{(Q)}&=&\left(\zeta_R^{(+1/2)},~\eta_R^{(-1/2)}\right)
\nn
Y&=&~~+1/2,~-1/2
\ee
While this case is a logical possibility from the point of view of the hypercharge assignment, we will see that it is disfavored by the precision measurements.  
\subsection{A more general  hypercharge assignment}
In general, all of the anomalies are avoided using the following 
generic hypercharge assignment:
\begin{eqnarray}
Y(T_L)& = &\frac{y}{2} \ ,\qquad Y(U_R,D_R)=\left(\frac{y+1}{2},\frac{y-1}{2}\right) \ , \\
Y({\cal L}_L)& = &-3\frac{y}{2} \ ,\qquad Y({\nu_{\zeta}}_R,\zeta_R)=\left(\frac{-3y+1}{2},\frac{-3y-1}{2}\right) \ .
\end{eqnarray}
One recovers the previous choices of the hypercharge for $y=1/3$ (standard model like family) and $y=0$ (fractionally charged leptons). Another choice of the hypercharge which does not lead to either fractionally charged techniquarks or leptons is, for example, $y=1$. In this case:
\begin{eqnarray}
Q(U)=1 \ , \quad Q(D)=0 \ , \quad Q(\nu_{\zeta})=-1 \ , \quad {\rm and }\quad Q(\zeta)=-2 \ , \quad {\rm with} \quad y=1 \ .
\end{eqnarray}
We will also analyze this possibility from the point of view of the electroweak precision measurements. In this case the dark matter candidate could be a neutral ditechniquark with the quantum numbers, for example, of 
two $D$ techniquarks. A more complete analysis is however needed along the lines suggested first in \cite{Chivukula:1989qb} for ordinary technicolor theories.

\subsection{The Technihadron Spectrum}
Besides the enlarged leptonic sector the technihadron spectrum is
also very interesting. {}Since the representations of two
colors are pseudoreal the global symmetry group is $SU(4)\supset
SU(2)_L\times SU(2)_R$, which is spontaneously broken to $O(4)$. 
This leads to nine Goldstone bosons of which three are eaten by the massive
standard model gauge bosons. 
The $U(1)$ of technibaryon number is contained in $SU(4)$ and 
can be identified with one of the diagonal generators of $SU(4)$ \footnote{For a complete set of generators of $SU(4)$ suitable 
for this case see the appendix of the second reference in \cite{Appelquist:1998xf}.}.
Hence some of these Goldstone bosons are
charged under the technibaryon number and are ditechniquarks.
Typically, these extra global symmetries are broken by ETC interactions.

We expect that the nature of the technibaryons, if observed, will help to understand the underlying gauge 
structure of the technicolor theory. Recall how the baryonic sector of ordinary QCD gave the first information 
on the nature and representation of the quarks.  

Here the low-lying technibaryons are made of two techniquarks. This can be understood by constructing the 
simplest examples of technibayons in the theory explicitly:
\begin{eqnarray}
B_{\{f_1,f_2\}} = T^{\{C_1,C_2\}}_{L;\alpha,
f_1}T^{\{ C_3,C_4 \}}_{L;\beta,f_2} \epsilon^{\alpha \beta}
\epsilon_{C_1 C_3}\epsilon_{C_2 C_4} \ .
\end{eqnarray}
where $f_i=1,2$ corresponds to $U$ and $D$ flavors and where we are symmetrizing in flavor. $\alpha$ and $\beta$ assume the values of one or 
two and represent the ordinary spin. These states  correspond to a $SU(2)_L$ triplet 
of scalar technibaryons. Similarly we can construct the technibaryons using only right fields. A technibaryon $SU(2)_L$ singlet can be constructed  but must have ordinary spin one. This is due to the fact that we 
also have to antisymmetrize with respect to the flavor indices. Note that we are not using the full $SU(4)$ flavor symmetry to classify the states. 

According to the underlying hypercharge assignment the technibaryon spectrum will lead to specific signatures at LHC and the detailed phenomenological consequences will be explored elsewhere. 

The scalar sector is also very
interesting since the neutral
Higgs can be very light with respect to
ordinary technicolor theories as we shall see.


\section{Three Technicolors: Minimal Modification of the Standard Model}
\label{4}

The theory with three technicolors contains an even number of electroweak doublets, and hence
it is not subject to a Witten anomaly.  
The doublet of technifermions, is then represented again as:
\be
T_L^{\{C_1,C_2 \}}
=
\left(\begin{array}{l}U^{\{C_1,C_2 \}}\\ D^{\{C_1,C_2 \}}\end{array}\right)_L \ ,
\qquad 
T_R^{\{C_1,C_2\}}&=&\left(U_R^{\{C_1,C_2\}},~ D_R^{\{C_1,C_2\}}\right) \ .
\nn
\ee
Here $C_i=1,2,3$ is the technicolor index and $T_{L(R)}$ is a doublet (singlet) with respect 
to the weak interactions. 
Since the two-index symmetric representation of $SU(3)$ is complex the flavor symmetry is $SU(2)_L\times SU(2)_R$. 
Only three Goldstones emerge and
are absorbed in the longitudinal components of the weak vector bosons.

Gauge anomalies are absent with the choice $Y=0$ for the hypercharge of the left-handed technifermions:
\be
T_L^{(Q)}
=
\left(\begin{array}{l} U^{(+1/2)}\\ D^{(-1/2)}\end{array}\right)_L \ .
\ee
Consistency requires for the right-handed technifermions (isospin singlets):
\be
T_R^{(Q)}&=&\left(U_R^{(+1/2)},~D_R^{-1/2}\right)
\nn
Y&=&~~+1/2,~-1/2 \ .
\ee 
All of these states will be bound into hadrons. There is no need for an associated fourth family of leptons, and hence it is not expected to be observed in the experiments.

Here the low-lying technibaryons are fermions constructed with three techniquarks in the following way:
\begin{eqnarray}
B_{f_1,f_2,f_3;\alpha} = T^{\{C_1,C_2 \}}_{L;\alpha,
f_1}T^{\{ C_3,C_4 \}}_{L;\beta,f_2} T^{\{ C_5,C_6 \}}_{L;\gamma,f_3}\epsilon^{\beta \gamma}
\epsilon_{C_1 C_3 C_5}\epsilon_{C_2 C_4 C_6} \ .
\end{eqnarray}
where $f_i=1,2$ corresponds to $U$ and $D$ flavors, and we are not specifying the flavor symmetrization which in any 
event will have to be such that the full technibaryon wave function is fully antisymmetrized in technicolor, flavor and spin. 
$\alpha$, $\beta$, and $\gamma$ assume the values of one or 
two and represent the ordinary spin. Similarly we can construct different technibaryons using only right fields or a mixture of left and right.
\section{General Constraints from Electroweak Precision Data}
\label{5}

In this section we confront our models with the electroweak precision measurements. The relevant corrections to the minimal standard model appear in the vacuum polarizations of the electroweak gauge bosons. These can be parameterized in terms of the three 
quantities $S$, $T$, and $U$ (the oblique parameters) 
\cite{Peskin:1990zt,Peskin:1991sw,Kennedy:1990ib,Altarelli:1990zd}, and confronted with the electroweak precision data. The relevant formulae we use from \cite{He:2001tp} and \cite{Gates:1991uu} to estimate these parameters are reported in the Appendix.

The models considered here produce smaller values of $S$ than
traditional technicolor models, because of their smaller particle
content and because of their near-conformal dynamics \cite{Hong:2004td}.

\subsection{Two technicolors and two techniflavors}
Different contributions must be taken into account when confronting this theory with the electroweak measurements. The first is the one arising due to the technicolor sector per se and the second is the one due to the new lepton family needed to guarantee that no anomalies are present in the theory. We will show that the theories proposed in \cite{Sannino:2004qp} have the smallest $S$ parameter when compared with other theories of walking technicolor \cite{Hong:2004td}, and the new precision measurements strongly favor non-degenerate leptons.  

We consider different ways of giving masses to the new lepton family, as described in the previous sections, and their effects on the electroweak precision measurements.

The oblique corrections due to the presence of the new lepton family can be readily computed both for a Dirac 
mass \cite{He:2001tp} as well as a combined Dirac and Majorana mass \cite{{Bertolini:1990ek},Gates:1991uu} for 
the leptons. {}For the estimates of the oblique corrections due to the strongly interacting sector 
non-perturbative corrections can play a role. {}Fortunately, for a walking technicolor theory we can use the perturbative results also for the techniquarks. This is so since for a theory of walking technicolor the perturbative corrections can be seen as conservative estimates. Indeed, it has been shown in \cite{{Sundrum:1991rf},Appelquist:1998xf} that the non-perturbative corrections reduce the perturbative contribution to the $S$ parameter. The reduction, roughly speaking,  is due to the fact that near the conformal fixed point chiral symmetry is expected to be almost restored. Since the $S$ parameter must vanish in a chirally symmetric theory we expect a further reduction to the perturbative value of the parameter $S$.  This is a universal feature which does not depend on the techniquark's representation. The reduction is of the order of $20\%$ \cite{Sundrum:1991rf}. 
Although, in general, a reduction is expected for the $S$ parameter, it is hard to see how a negative $S$ 
parameter can emerge in a walking technicolor theory only due to non-perturbative 
effects and without considering large weak isospin violations. Effects on the oblique parameters due to the energy dependent behavior of the masses of fermions in the 
standard model generated by the dynamics of the walking type theory 
have also been investigated recently in \cite{Christensen:2005hm}.

The simple perturbative estimate leads to the following value of the $S$ parameter:
\begin{equation}
{ S}_{\rm pert.} = {1 \over 2 \pi}  \ .
\end{equation}
 This is the walking technicolor theory with the smallest perturbative $S$ parameter \cite{Hong:2004td}. 
However one cannot compare this result immediately with the electroweak precision data since this theory has a Witten anomaly and hence the new lepton family must be included in the analysis when comparing with the precision measurements.

\subsection{Testing our model including the contribution of the Fourth Lepton Family}

Here we consider the combined effect on the electroweak precision measurements of this walking technicolor theory 
together with the fourth family of leptons. If we choose the hypercharge for the leptons to be the one of ordinary leptons one can easily 
see that from the electroweak point of view the theory features effectively a complete fourth family of leptons
 and (techni)quarks. This is so since the techniquark doublet, being in the two-index symmetric representation of the $SU(2)$ gauge group, comes exactly in three copies with respect to the electroweak interactions. 
 
\subsubsection{Dirac Masses for the Technifermions and the fourth family of Leptons}
A new fourth family with mass degenerate fermions is ruled out at more than $90\%$ level of confidence since 
the associated $S$ parameter would be positive and too large. Non-degenerate heavy Dirac 
fermion doublets can substantially decrease the 
value of $S$ at the expenses of a non-zero and always positive value of the $T$ parameter. We show that for 
the present theory, a small splitting of the fermion masses is sufficient to make the model an economical 
and elegant candidate for a mechanism dynamically breaking the symmetry of the electroweak theory.
 
\begin{figure}[htbp]
  \begin{center}
    \mbox{
      \subfigure[Perturbative Techniquarks]{\resizebox{!}{4.5cm}{\includegraphics{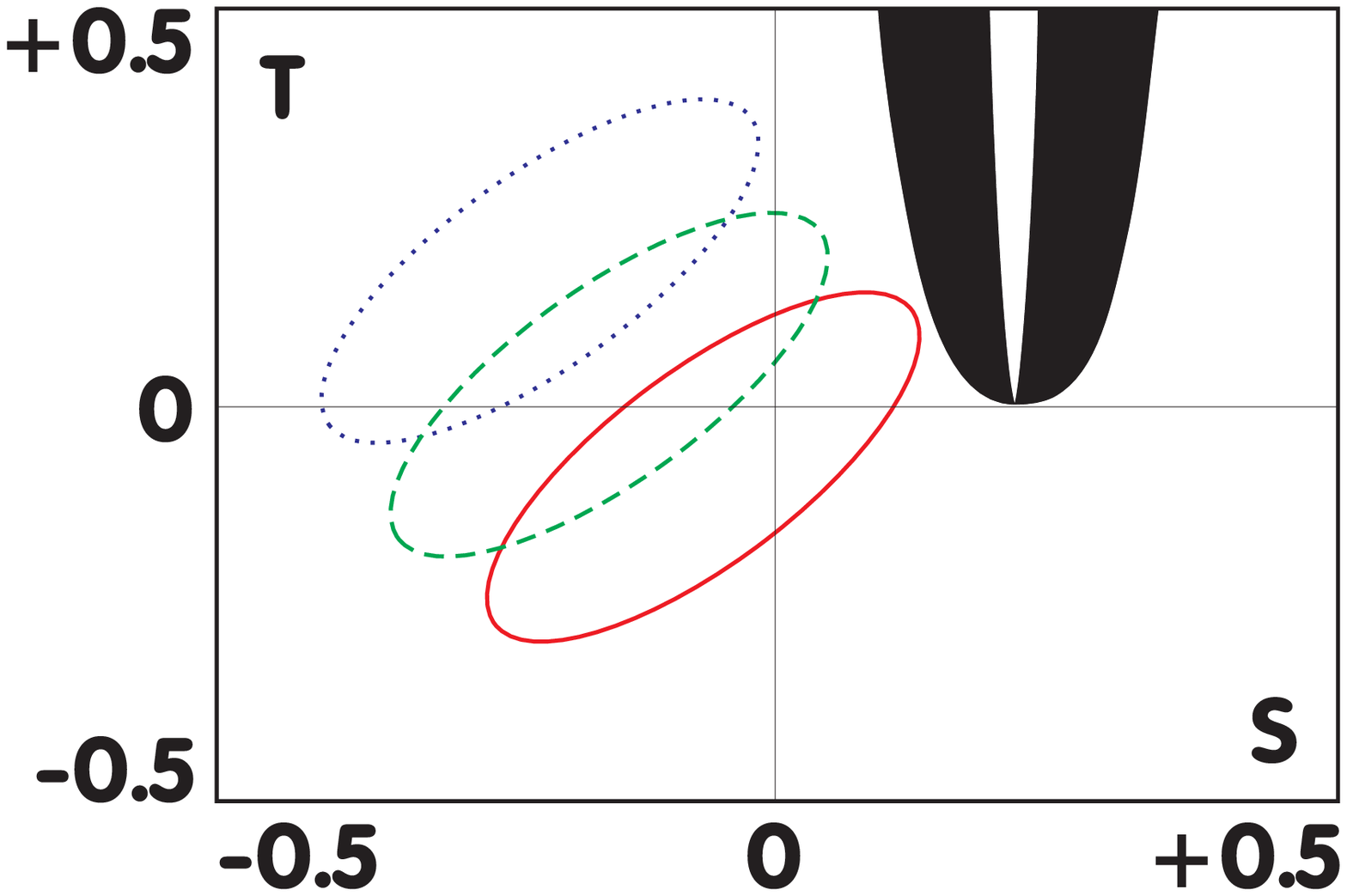}}} \qquad 
      \subfigure[Nonperturbative Techniquarks]{\resizebox{!}{4.5cm}{\includegraphics{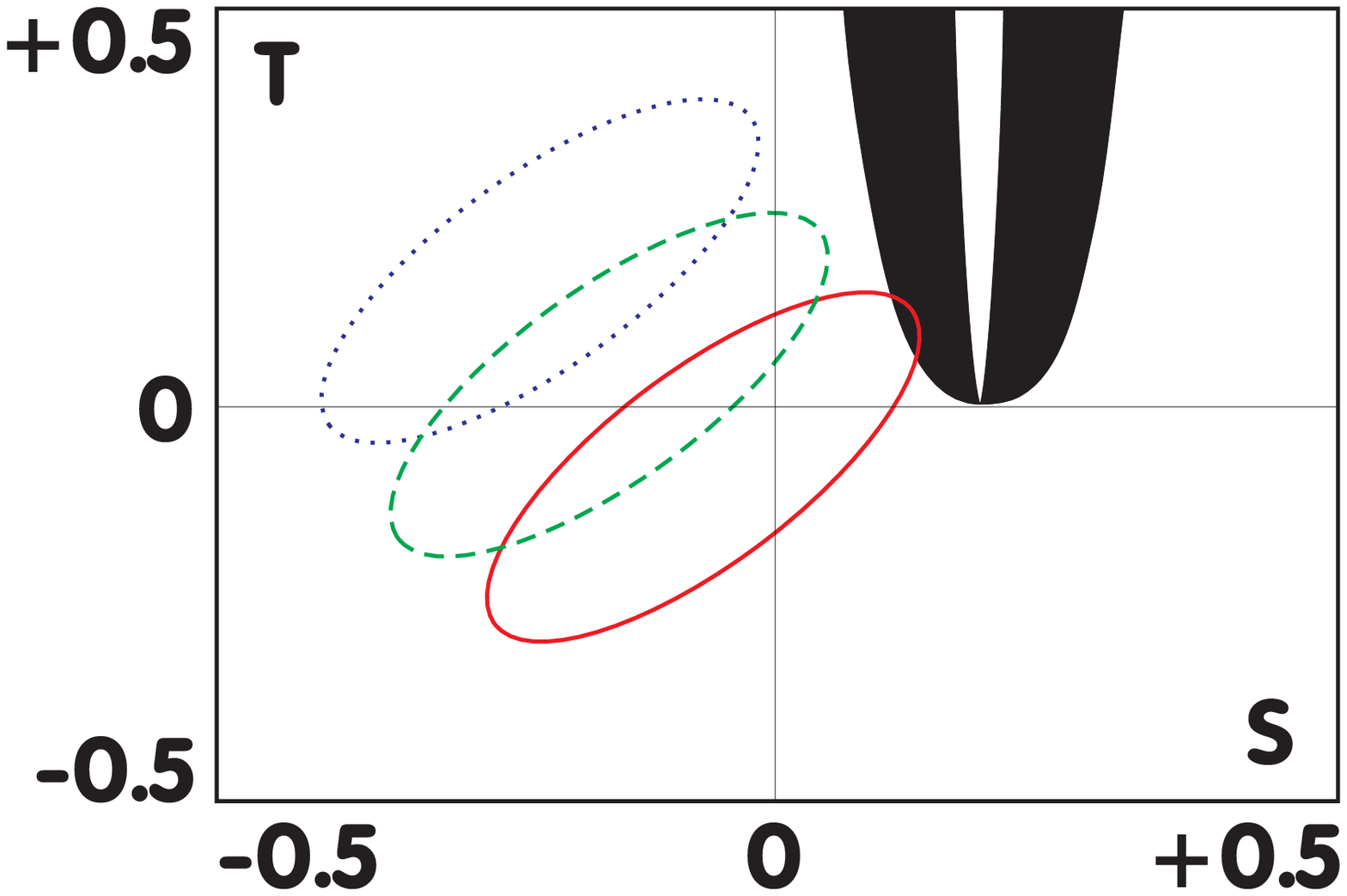}}} 
      }
    \caption{Left Panel: The black shaded parabolic area corresponds to the accessible range of $S$ and $T$ for 
    the extra neutrino and extra electron for masses from $m_Z$ to $10 m_Z$. The perturbative estimate for the 
    contribution to $S$ from techniquarks equals $1/2\pi$. The ellipses are the $90$\% confidence level contours 
for the global fit to the electroweak precision data with $U$ kept at $0$. The values of $U$ in our model lie 
typically between $0$ and $0.05$ whence they are consistent with these contours. The contours from bottom to top 
are for Higgs masses of $m_H = 117$, $340$, $1000$ GeV, respectively. Right Panel: We added non-perturbative 
corrections to the $S$ parameter in the technicolor sector of the theory.}
    \label{fig:ST0}
  \end{center}
\end{figure}

In figure \ref{fig:ST0} we show the results allowing for mass splitting for the leptons but keeping degenerate
 techniquarks. 
In the left panel we present the accessible range of the two oblique parameters $S$ and $T$ when the extra 
neutrino and electron masses range from $m_Z$ to $10~m_Z$. In the left panel we consider the perturbative value 
$1/2\pi$ for the contribution from the techniquarks. The ellipses are the $90\%$ confidence level contours for 
the global fit to the electroweak precision data to be found in the latest review of the Review of Particle 
Properties \cite{Eidelman:2004wy} with $U$ kept at $0$. The values of $U$ in our model lie typically between $0$ 
and $0.05$ whence they are consistent with these contours. The contours from bottom to top are for Higgs masses 
of  $m_H = 117$, $340$, $1000$ GeV, respectively. 

There are also non-perturbative corrections which further reduce the techniquark contribution to the $S$ parameter, whereby they bring the theory even 
closer to the precision measurements as can be seen from the right panel of figure \ref{fig:ST0}. 
{}To be more precise, near-conformal dynamics leads to a further reduction in
the $S$ parameter \cite{Sundrum:1991rf,Appelquist:1998xf}.
In the estimate of \cite{Sundrum:1991rf},
based
on the operator product expansion, the factor of ${1 \over 6 \pi}$
in the expression for the perturbative value of $S_{pert}=\frac{1}{6\pi} N(N+1)/2$ for one doublet of technifermions in the 
two-index symmetric representation of the $SU(N)$ gauge theory  is reduced to about $.04$, which is roughly a 
twenty percent reduction. This correction has been used to produce the figure in the right panel of figure \ref{fig:ST0}.

Note that current models of walking type with fermions in the fundamental representation are disfavored 
by the data. This is clear already when considering the perturbative as well as non-perturbative computation of the associated $S$ parameter \cite{Hong:2004td}.

\subsubsection{Dirac \& Majorana Masses}
We have also considered the possibility that the new lepton family has a Majorana mass \`a la Bertolini and Sirlin \cite{Bertolini:1990ek}, 
which also seems to provide a negative contribution to the $S$ parameter \cite{Gates:1991uu}. However, when investigating the range of attainable points in the S-T-plane the combination of the Majorana with the Dirac mass does not lead to a larger overlap with the $90\%$ level of confidence contours. Nevertheless, this democratic choice is still a possible extension beyond the simple Dirac case. 

\subsubsection{Zero Hypercharge: Fractionally Charged Lepton}
The model with fractionally charged leptons can hardly improve the $S$ parameter problem. Indeed, its predictions for the $S$ and $T$  
parameter is basically on a vertical line centered in the gap of the inner parabola presented in the previous figures. This is so 
since the fractionally charged leptons have an associated zero hypercharge.

\subsubsection{Higher Hypercharge with a Doubly Charged Lepton}
\begin{figure}[htbp]
  \begin{center}
    \mbox{
      \subfigure[Perturbative Techniquarks]{\resizebox{!}{4.5cm}{\includegraphics{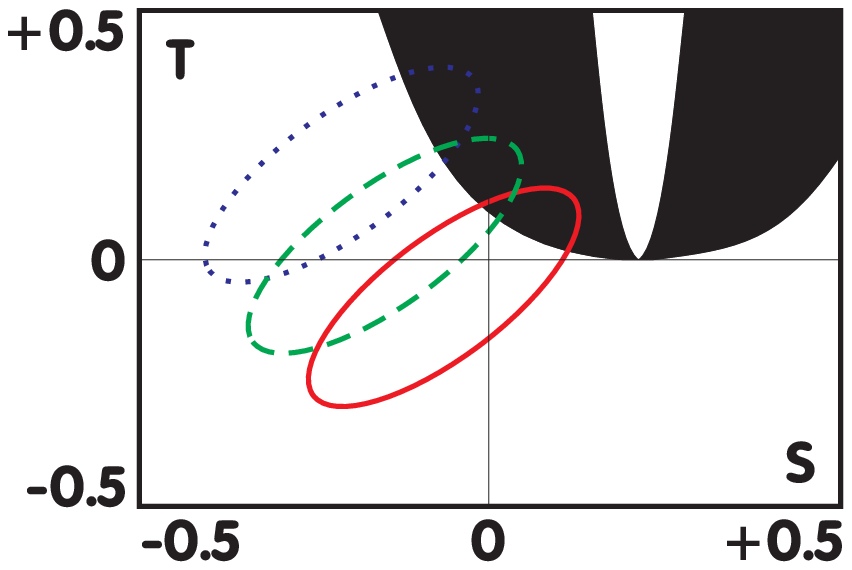}}} \qquad 
      \subfigure[Nonperturbative Techniquarks]{\resizebox{!}{4.5cm}{\includegraphics{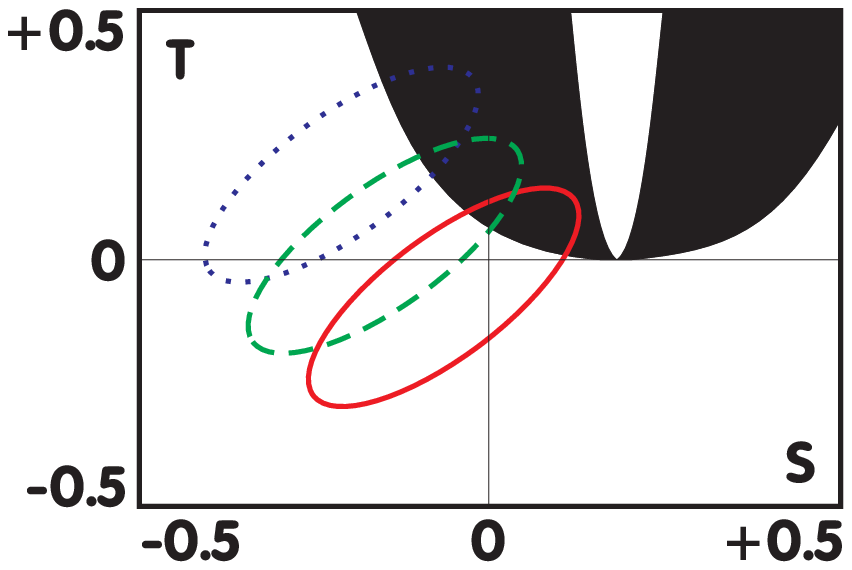}}} 
}
    \caption{Left Panel: The black shaded parabolic area corresponds to the accessible range of $S$ and $T$ for 
    the new singly and doubly charged leptons with masses from $m_Z$ to $10 m_Z$. The perturbative estimate for the 
    contribution to S from the techniquarks is $1/2\pi$. The ellipses are the $90$\% confidence level contours 
    for the global fit to the electroweak precision data with $U$ kept at $0$. The values of $U$ in our model lie 
    typically between $0$ and $0.05$ and hence they are consistent with these contours. The contours from bottom 
    to top are for Higgs masses of  $m_H = 117$, $340$, $1000$ GeV, respectively. Right Panel: We added non-perturbative corrections to the $S$ parameter in the
technicolor sector of the theory.}
    \label{fig:STY}
  \end{center}
\end{figure}

We have also considered the case in which all of the new fermions have integer electric charges. In our example we have chosen the lowest possible integer charge for the fermions compatible with the absence of anomalies. The new lepton family features a doubly as well as a singly charged lepton. Interestingly, when comparing the predictions for the oblique corrections due to the model with the data we find a larger overlap with the confidence level contours (see figure \ref{fig:STY}). Such an improvement is due to the fact that the hypercharge for the new leptons is larger than in the cases investigated above, while the technicolor sector contribution is unaffected in the limit of degenerate techniquarks. 

In reference \cite{Peskin:2001rw} Peskin and Wells already lined out different ways, elaborated in the past decade, to save traditional models of dynamical breaking of the electroweak sector from being ruled out by the data from precision measurements. Our results 
explicitly show that the method of positive $T$ \cite{Peskin:2001rw} is sufficient to bring the composite 
Higgs theory within the experimental acceptable range when the technifermions are in higher representations of the technicolor gauge group. 
{A heavy fourth family of ordinary quarks and/or leptons has been investigated theoretically and experimentally in the past, see \cite{Eidelman:2004wy} for an up to date review and also \cite{Froggatt:1996um,{Froggatt:2002vs}}}. 

\subsection{Contribution from Technicolor with three colors and two flavors}
We have also investigated the case of three technicolors and two techniflavors. This theory has no Witten or gauge anomalies and hence 
no new lepton family is needed. When computing the perturbative $S$ parameter one discovers that it is larger than the one for the 
two technicolor theory, actually twice as large, and hence it 
is harder to make it fit with the precision data especially without the inclusion of additional lepton families which are not a must as in the case of 
two colors.

\section{Lepton Spectrum at LHC and a Dark Matter candidate? }
\label{6}

The general feature of the two technicolor theory with technifermions in the two-index symmetric 
representation of the gauge group is 
the necessity to include at least one new lepton family.
If the theory underlying the spontaneous breaking of the electroweak symmetry is of the type 
presented above, the current precision measurements are already sufficient to allow us to make 
specific predictions on the new associated leptonic sector. 
This might also guide, in the future, the construction of extended technicolor models.

Assigning standard model like charges for the three 
techniquarks requires also standard model type charges for the fourth family of leptons. 

\begin{figure}[htbp]
\begin{center}
    \includegraphics[width=5truecm,height=5truecm]{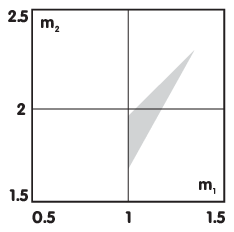}
     
Charged vs Neutral Lepton Mass Spectrum
    \caption{This plot summarizes the spectrum of the new fourth family of leptons with $m_1$($m_2$) the new neutrino(lepton) mass in units of $m_Z$ for a standard model like hypercharge assignment allowed by 
electroweak precision measurements at the $90\%$ level of confidence. We kept the non-perturbative corrections for the computation of the $S$ parameter for the technicolor contribution. The polygonal shape corresponds to the overlap of the parabolic shape with the interior of the rectangular contour defined by $S<0.12$ and $T<0.13$ approximating the ellipse for $m_H=117$~GeV.}
    \label{fig:masslept}
  \end{center}
\end{figure}
{}From the plot it is clear that the neutral lepton is always lighter than the charged one. 
Interestingly we predict that at LHC one might discover a fourth family of ordinary leptons while the associated quarks would be bound into objects which do not interact strongly but are the technihadrons associated  to the electroweak theory. The Higgs must be light and this is consistent with our estimates provided in the next section.  

The new neutrino has a mass between $m_Z$ and $1.5~m_Z$ (see Fig. \ref{fig:masslept}), while the associated negatively charged lepton has a mass of roughly twice the mass of the neutral weak gauge boson. Interestingly the new neutrino, if made stable, could be a natural cold dark matter candidate.

\begin{figure}[htbp]
  \begin{center}
    \mbox{
      \subfigure[]{\resizebox{!}{4cm}{\includegraphics{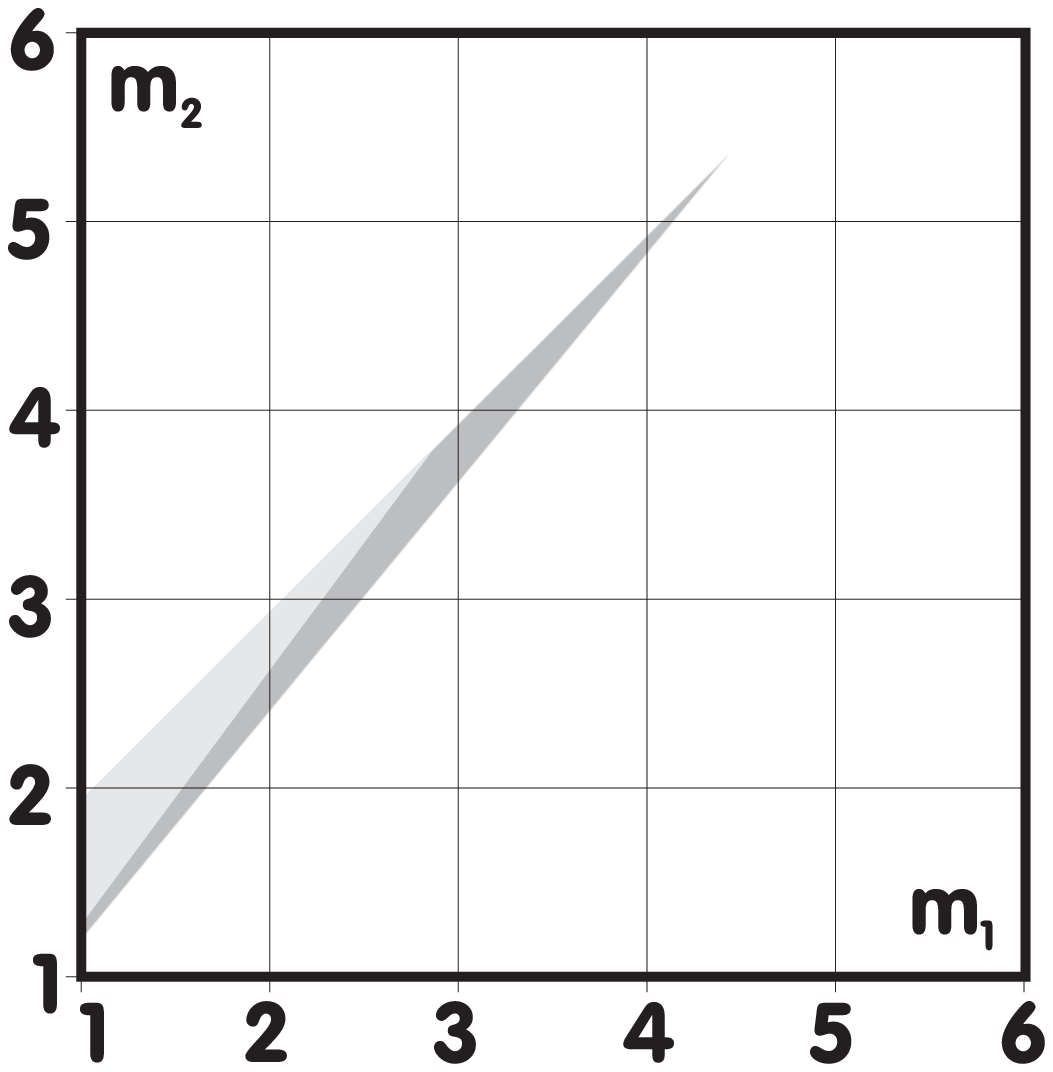}}} \qquad 
      \subfigure[]{\resizebox{!}{4cm}{\includegraphics{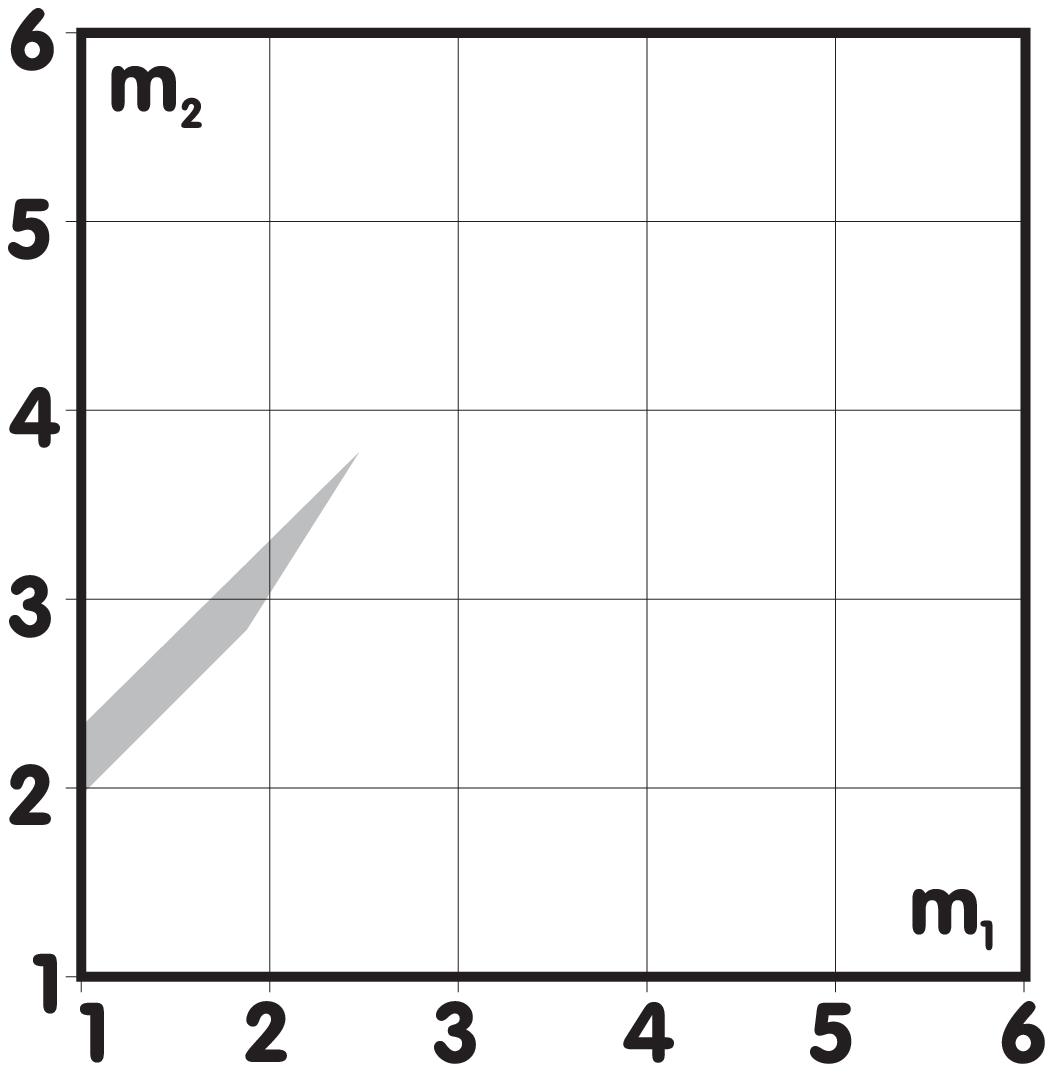}}}\qquad 
      \subfigure[]{\resizebox{!}{4cm}{\includegraphics{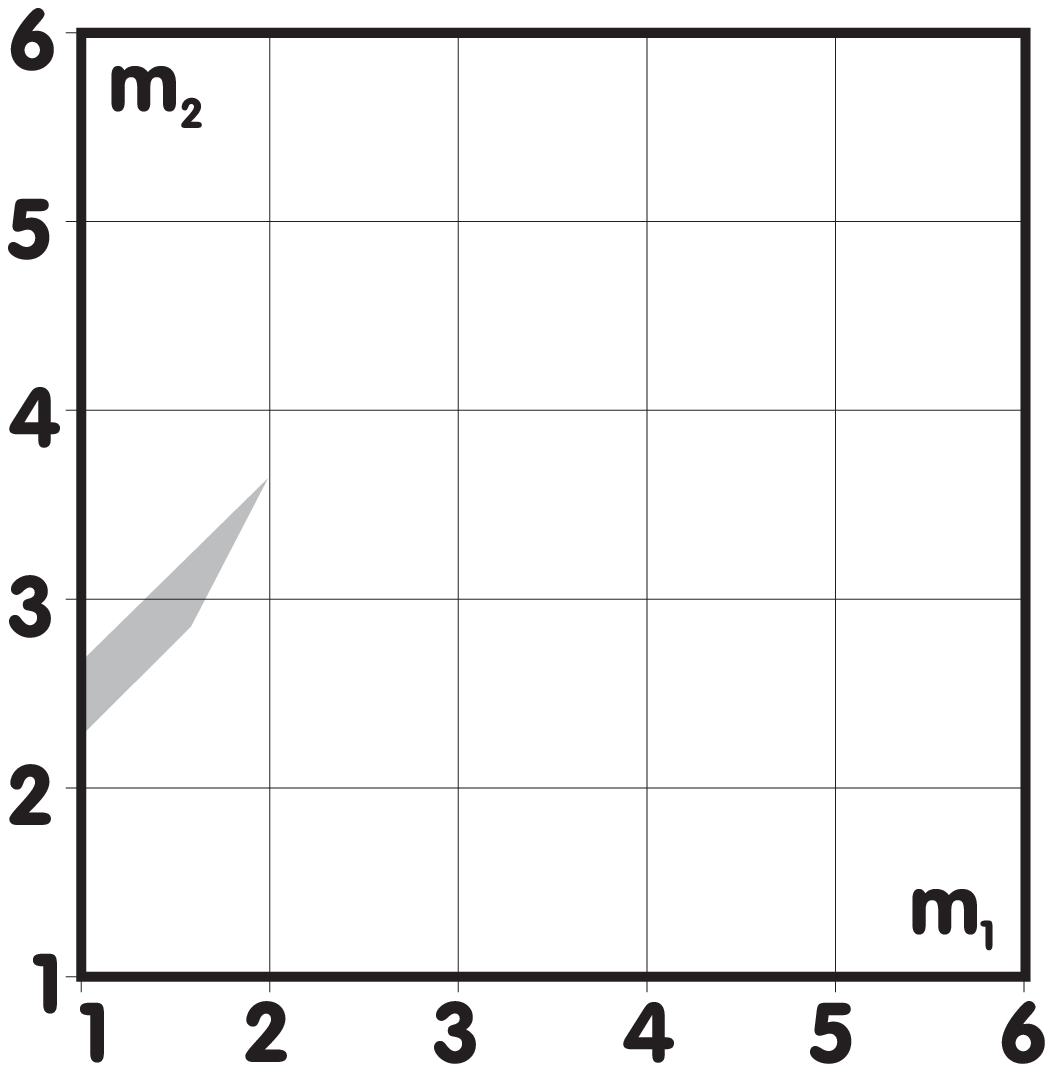}}} 
      }
    \caption{The shaded areas depict the accessible range of the new lepton masses 
    due to the oblique corrections. $m_1$($m_2$) is the mass parameter in units of $m_Z$ of the singly(doubly) charged lepton. The polygonal shapes do not correspond exactly to the ellipses in the S-T-plane but to rectangular areas defined by:
a) $S<0.12$ and $T<0.13$ for $m_H=117$~GeV;  
b) $S<0.04$ and $0.13<T<0.24$ for $m_H=340$~GeV; and c) $S<-0.02$ and $0.23<T<0.38$ for $m_H=1$~TeV. The shaded  areas in all of the figures correspond to the non-perturbative evaluation of the technicolor contribution to the $S$-parameter. In the plot for $m_H=117$~GeV the light grey part corresponds to the purely perturbative evaluation.}
    \label{fig:DCL}
  \end{center}
\end{figure}

We now turn to the case of a higher hypercharge assignment featuring a singly and a doubly (electrically) charged 
lepton. In figure \ref{fig:DCL} we show the accessible range of the lepton masses. 
Since we expect a very light Higgs, figure (a) is the relevant one. We then predict
 the doubly charged lepton to be heavier than the associated singly charged one. It is unstable and 
 decays into the associated singly charged lepton. The experimental bounds on the existence and properties of a doubly charged lepton are very weak \cite{Eidelman:2004wy}. 
\section{Light Higgs from Higher Representations}
\label{7}

In extensions of the standard model featuring an elementary Higgs boson, a very strong constraint 
on its mass can be set by requiring that the interactions, in Nature, are weakly-coupled up 
to a grand unification scale of the order of $10^{16}$GeV \cite{Cabibbo:1979ay}. {}For example in generic supersymmetric extensions of the standard model addressing the related hierarchy  one usually obtains an upper 
bound of $205$~GeV \cite{Kane:1992kq,{Espinosa:1992hp},{Espinosa:1998re}}. {It is interesting to observe that there are also 
 principles which are suited for an elementary Higgs boson and that have been used to make a number of phenomenological predictions \cite{Froggatt:1995rt}. }

It would then be very interesting to provide examples of composite Higgs theories featuring 
a light Higgs boson which is not a quasi Goldstone boson.

In the analysis of QCD-like technicolor models information on the non-perturbative dynamics at the electroweak scale is obtained by simply scaling
up QCD phenomenology to the electroweak
energy scale. The Higgs particle is then mapped into the scalar chiral-partner of the Goldstone bosons in QCD. 
The scalars represent 
a very interesting and complicated sector of QCD. Much work has been devoted to providing a 
 better understanding of this sector which is relevant to understand the vacuum structure 
 of QCD. There is a growing consensus that the low lying scalar object, i.e. $f_0(600)$,  needed to provide a good description of low energy pion pion scattering \cite{Sannino:1995ik} 
 is not the chiral partner of the pions but is of four quark nature \`a la Jaffe \cite{Jaffe:1976ig,{Jaffe:1976ih}}.

Recent arguments, based on taking the limit of a large number of colors $N$, also demonstrate that the low energy scalar is not of $q\bar{q}$ nature \cite{Harada:2003em,Pelaez:2003dy,Uehara:2003ax}. The natural candidate for the chiral partner of the ordinary pions is then very heavy, i.e. it has a mass larger than one GeV, and this experimental result agrees with naive scaling estimates. When transposed to the electroweak theory by simply taking $F_{\pi}$ as the electroweak scale, one concludes that in technicolor theories with QCD-type dynamics the Higgs is very heavy, $m_H\sim 4 \pi F_{\pi}$, of the order of the TeV scale. This also means that large corrections are needed, due to new physics, to 
compensate the effects of such a heavy Higgs with respect to the electroweak precision measurements data.

{}However, for strongly interacting theories with non-QCD-like dynamics we are no longer guaranteed that the 
dynamically generated Higgs particle is heavy
\footnote{Here we are not considering theories in
which the Higgs is a quasi Goldstone boson of some strongly
interacting theory, i.e. the so called little Higgs theories.}.
In particular the QCD-like estimates cannot be
trusted in walking technicolor or other near-conformal models. This is especially true for the theories with fermions 
in higher representation of the gauge group. One
cannot simply scale up QCD to obtain useful non-pertubative
information. 

One of the main problems when considering non-QCD-like theories to construct possible extensions of the standard 
model is the lack of specific predictions on the non-perturbative dynamics. 
Very recently we have shown \cite{Hong:2004td,{Sannino:2004qp}} that it is possible 
to provide new information on the hadronic sector (and therefore the Higgs mass) of theories with higher representation by studying the one flavor sector of the S(A)-type theories. 
This was possible due to the recent observation \cite{Armoni:2004uu} that {\em
non}-supersymmetric Yang-Mills theories with a Dirac fermion
either in the two-index symmetric or antisymmetric representation
of the gauge group are non-perturbatively equivalent to
supersymmetric Yang-Mills theory (SYM) at large $N$. This allowed to export some exact
results established in SYM to the non-supersymmetric theories. These 
theories are also called 
``orientifold'' theories. At finite $N$ they were studied in \cite{Sannino:2003xe} and many of the discovered
properties, such as almost exact parity doubling and small vacuum
energy density, are appealing properties for dynamical breaking of
the electroweak theory \cite{Appelquist:1998xf}. We emphasize
again that supersymmetry is only used as a technical tool here to
extract non-perturbative information about our models, which are
not themselves supersymmetric. It is also worth recalling that technicolor theories with fermions 
in the two-index symmetric representation of the gauge group were dubbed by two of the present authors ``techniorientifold'' theories in \cite{Sannino:2004qp} and soon after were referred to as composite Higgs theories 
from higher dimensional representations \cite{Hong:2004td}. We have kept in this paper the latter, more intuitive terminology. {}In this section and for the reader's convenience 
we will first summarize how to determine qualitatively the mass of the lowest lying fermion-antifermion scalar meson with respect to the invariant scale of the theory in the one flavor theory of S(A)-type. This shows that for the S-type theory this object is substantially lighter than in ordinary QCD \cite{Hong:2004td}. 

Then we will propose a simple and new method which will allow us to extrapolate the mass of the scalar meson, associated to chiral symmetry breaking, to the case of 
two (techni)flavors. This method uses the fact that we are near a chiral phase transition as well as a conformal fixed point. These 
two facts insure that the chiral partner of the Goldstone excitations must become light at the phase transition.
 
Using the saturation of the underlying trace anomaly via the effective theory built out of the scalar meson we will show that near the phase transition the mass squared of the scalar field 
depends linearly on the difference between the critical value of the number of (techni)flavors and the generic number of (techni)flavors, i.e.:
\begin{eqnarray}
M^2_{\sigma} \propto N^c_{T f} - N_{T f} \ .
\end{eqnarray} 
This is the standard mean field theory behavior of the squared mass associated to the order parameter field with respect to the scaling field. We will discuss the possible problems associated with this picture and the current status in the literature. Note that the present phase transition is an example of a  quantum phase transition \cite{Shopova:2003jj,{Vojta},{Sachdev}}. 

If this near conformal and chiral dynamics is used to model electroweak symmetry breaking the previous fermion-antifermion scalar (the $\sigma$) becomes automatically the Higgs of the theory. Since we 
can provide an estimate for the mass of this scalar for one flavor we can extrapolate it for the physically interesting case of 
two flavors and show that the Higgs particle is very light in theories with underlying fermions in the two-index 
symmetric representation of the gauge group. Our method is general and we can also estimate the Higgs mass within ordinary walking technicolor theories in which the fermions are in the fundamental representation of the gauge group. Here the Higgs is also lighter than in QCD-type theories but still heavier than in near conformal S-type theories.

\subsection{The $N_{Tf}=1$ sector via Supersymmetry}

We review here how to estimate the mass for the low lying scalar meson in the S-type theory for
$N_{Tf}=1$, first derived in \cite{Hong:2004td} in the large $N$ limit.
This scalar is first identified with the associated
fermion-antifermion state whose pseudoscalar partner in ordinary
QCD is the $\eta^{\prime}$. At large $N$ this theory is mapped into the bosonic sector of super 
Yang-Mills \cite{Armoni:2004uu}. The low lying
bosonic sector contains precisely a scalar and a pseudoscalar
meson. Due to the holomorphicity property of supersymmetry the two bosons become degenerate at infinite $N$. 
Besides, in the
supersymmetric limit we can relate the masses to the fermion
condensate $\left< \bar{T}_R\,T_L \right> \equiv \left<
\bar{T}_R^{\{i,j \}} {T}_{L,\{i,j \}} \right> $
\cite{Sannino:2003xe}:
\begin{eqnarray}
M=\frac{2\,\alpha}{3}\, \left[\frac{3 \left< \bar{T}_R\,T_L
\right> }{32\pi^2\,N} \right]^{\frac{1}{3}} =
\frac{2\hat{\alpha}}{3} \Lambda \ ,
\end{eqnarray}
with $\left< \bar{T}_R\, T_L \right> = 3N\Lambda^3$ and $\Lambda$
the one loop, and large $N$ invariant scale of the theory:
\begin{eqnarray}
\Lambda^3 = \mu^3 \left(\frac{16\pi^2}{3Ng^2(\mu)}\right) \exp
\left[ \frac{-8\pi^2}{Ng^2(\mu^2)}\right] \ .
\end{eqnarray}
We have also defined:
\begin{eqnarray}
\hat{\alpha} = \alpha\, \left[\frac{9}{32\pi^2}
\right]^{\frac{1}{3}}\ .
\end{eqnarray}
The unknown numerical parameter $\hat{\alpha}$ is expected to be of
order one, and is the coefficient of the K\"{a}hler term in the
Veneziano-Yankielowicz effective Lagrangian describing the lowest
composite chiral superfield. Taking for example $\hat{\alpha} \sim
1-3$ (see the discussion below) one would roughly deduce, at large
$N$ and for $N_{T f}=1$, a scalar mass in the range:
\begin{eqnarray}
M_{\sigma}=M\simeq 200-500~{\rm GeV} \ .
\end{eqnarray}
Here we have chosen $\Lambda =\Lambda_{TC}\sim 250~$GeV
\footnote{Strictly speaking, $F_{TC} = 250$ GeV. The relation
between $F_{TC}$ and $\Lambda_{TC}$ is given for two-index matter
(at large $N$) by $F_{TC} = c N \Lambda$. We took $c N$ of order
one; in QCD the large $N$ relation $F_{\pi} = c' \sqrt{N}
\Lambda_{QCD}$ implies $c'$ somewhat smaller than unity for
$F_{\pi} \sim 100$ MeV and $\Lambda_{QCD} \sim 300$ MeV, which is
consistent with $c N$ of order one.}. 
Clearly this cannot be directly the Higgs mass since we 
are considering the one flavor sector of the theory, however we will see that this mass is lighter 
when considering the $N_{T f}$ dependence. Besides, we expect $1/N$ corrections
to be important, and fortunately these corrections were estimated,
for $N_{T f}=1$, in \cite{Sannino:2003xe}. The corrections are
different for theories of type S and A. {}We
have:
\begin{eqnarray}
\frac{M_{\sigma}(S/A)}{M}=1 \mp \frac{4}{9N} + \frac{1}{8N}\frac{\langle
G_{\mu\nu}^a G^{a\mu\nu}\rangle}{\hat{\alpha}\,\Lambda^4} + {
O}(N^{-2}) \ ,
\end{eqnarray}
where $\langle G_{\mu\nu}^a G^{a\mu\nu}\rangle$ is the technigluon
condensate. The upper(lower) sign, in the previous equation, corresponds to the S(A)-type theory. 
Since $\langle G_{\mu\nu}^a G^{a\mu\nu}\rangle \sim
\Lambda^4$ and $\hat{\alpha}$ is order one the second term
dominates, and further reduces(increases) the scalar mass with respect to the
large $N$ limit in the S(A) theory. Since for $N=3$ the fermions, for type A
theories, are in the fundamental representation, our results are
qualitatively in agreement with the standard expectations that the
Higgs for theories with technifermions in the fundamental representation
is expected to be heavy.

To reassure ourselves that $\hat{\alpha}$ is indeed an order one
quantity we recall that the A-type theory with one flavor is mapped
into SYM at large $N$. {}But for $N=3$ the A-type
theory is QCD with one flavor, since the fundamental and two-index
antisymmetric representations are the same in $SU(3)$. This
observation was made long ago by Corrigan and Ramond
\cite{Corrigan:1979xf}. The $\eta^{\prime}$ state is the
pseudoscalar partner of the scalar fermion-antifermion state 
and its mass for one flavor can be simply estimated as
follows:
\begin{eqnarray}
M^2_{\eta^{\prime}}(N_{T f}=1) = \frac{N_{T f}}{3} M^2_{\eta^{\prime}} \ ,
\end{eqnarray}
where we used Witten and Veneziano's standard large $N$ and finite
$N_{T f}$ scaling, with $N=3$. Comparing this mass with the
supersymmetric limit by identifying $M$ with $M_{\eta^{\prime}}$ we
estimate:
\begin{eqnarray}
\hat{\alpha} \sim \frac{\sqrt{3}}{2} \frac{M_{\eta^{\prime}}}{
\Lambda} \sim 3.2\, \frac{\sqrt{3}}{2} \sim 2.8\ .
\end{eqnarray}
Here $M_{\eta^{\prime}}=958~$MeV is the ordinary 3-flavor QCD mass
for the $\eta^{\prime}$ and $\Lambda\sim 300~$MeV is identified
with the characteristic QCD invariant scale. It is
encouraging that we obtained the
suggested order one result used earlier. Lattice
simulations should be able to improve the estimate for
this mass.

\subsection{Near Quantum Phase Transition Flavor Dependence}

In the previous subsection we have studied the case of a single flavor and have shown that in theories 
with the two-index symmetric representation the lowest scalar fermion-antifermion state is lighter than in 
QCD-like theories \cite{Hong:2004td}. Here we consider the effects of including the flavor dependence. 
The method we adopt to determine such a dependence is applicable only if we have a nearby chiral phase transition as function of the number 
of flavors, for a fixed number of colors.  

Generally, near the critical point of a continuous phase
transition, the mass squared of the scalar order parameter drops
proportional to $(t-t_c)^{\nu}$, with $t$ the parameter driving
the phase transition, $t_c$ its value at the transition point, and
$\nu$ the critical exponent. One well-known example is ordinary
massless QCD near the chiral symmetry restoration point at finite
temperature. In this case the scalar partner of the pions must
become light close to the phase transition. So, despite its large mass
in vacuum, the scalar meson becomes very light near the phase
transition. Continuous phase transitions asymptotically close to the critical point 
display a
classical behavior. In other words, close to the transition, the thermal
fluctuations override the quantum ones \cite{Shopova:2003jj,{Vojta},{Sachdev}}
\footnote{More precisely one has a quantum phase transition when the de Broglie thermal wavelength $\lambda$ 
is greater than the correlation length of the thermal fluctuations $\xi$, i.e. $\lambda/\xi > 1$.}. {}For zero temperature field
theory which we are discussing here, the quantum effects remain
important. Such non-thermal phase transitions, dubbed quantum
phase transitions, have recently attracted much attention within
the condensed matter physics community. One well-known
example being the zero-temperature Bose-Einstein condensation. New applications of the Bose-Einstein 
condensation mechanism for the electroweak theory were presented in \cite{Sannino:2003mt,{Sannino:2003ai}} 
and also investigated recently in \cite{Loewe:2004zw} following the seminal work in \cite{Linde:1979pr,{Kapusta:1990qc},{Ferrer:1987jc},{Ferrer:1987ag}} . Interesting supersymmetric and ADS/CFT 
applications have also recently been proposed in \cite{{Harnik:2003ke},Apreda:2005yz}.

In the context of non-supersymmetric gauge theories the zero temperature chiral phase
transitions as function of the number of flavors have been studied using non-perturbative methods, e.g. in 
\cite{Appelquist:1991kn}. Here the authors already stressed that 
the phase transition as function of the number of flavors is very different from the one driven by temperature. 
Other interesting theories displaying quantum phase transitions as function of the number of flavors, although without 
gauge interactions, were investigated in \cite{Hamidian:1995pf}. Note that 
supersymmetric gauge theories also display phase transitions as function of the number of flavors for fixed number of colors. These transitions have been investigated in recent years \cite{Intriligator:1995au} and can also 
be understood as examples of quantum phase transitions. Actually 
one could reinterpret some of Seiberg's dual theories near the conformal phase transition as the effective theory for 
the quantum phase transition which can be used to compute the critical exponents and hence 
characterize their universal behavior. In this case the critical exponents of the theory would be just the ordinary four dimensional anomalous 
dimensions computed using the weakly coupled dual theory.

We approach the conformal window from the hadronic side of the theory by adjusting the number
of (techni)flavors with respect to the number of (techni)colors. So we treat the number of flavors $N_{Tf}$ as a
tunable scaling field. $N_{Tf}^c$ is the critical number of
flavors at which chiral symmetry is restored together with the onset of conformal invariance signaled by the vanishing of the 
underlying trace anomaly. We then construct an effective mean field theory and assume the scalar fermion-antifermion field to be the relevant one. 
The generic dependence of the squared mass on the number of flavors is:
\begin{eqnarray}
M^2_{\sigma}(N_{Tf})= \Lambda^2 (N_{Tf}^c - N_{Tf})^{\nu} \ , 
\end{eqnarray}
with $\nu$ a positive exponent and $\Lambda$ a flavor independent scale in the theory. Such a flavor independent 
scale in the theory can be eliminated by writing the previous expression as follows:
\begin{eqnarray}
M_{\sigma}^2(N_{Tf})=M_\sigma^2(\bar{N}_{Tf})\left[\frac{N_{Tf}^c-N_{Tf}}{N_{Tf}^c-\bar{N}_{Tf}}\right]^{\nu} \ ,
\end{eqnarray}
where $\bar{N}_{Tf}$ is some fixed value of the number of flavors sufficiently close to the critical value. Note that this flavor dependence is not what one obtains by computing the traces over the 
flavor space for the effective theory \cite{DiVecchia:1980xq,{DiVecchia:1980ve}}. Very near the critical point one can trade this overall flavor dependence for the constant value of the critical number of flavors. 

Due to the four dimensional nature of the present quantum phase transition fermions can no longer be neglected
near the phase transition \cite{Appelquist:1991kn,{Vojta}}. In a thermal phase transition the antiperiodic boundary conditions remove the fermionic zero modes from the low energy spectrum, while in a truly four dimensional phase transition such zero modes are present and are known to affect the universal behavior of these theories. This means that if we are interested in computing the critical exponents of the theory one should include from the hadronic side the possible composite fermions while from the nonhadronic side one should use the underlying gauge theory.

In this paper we are interested in the behavior of the lightest scalar fermion-antifermion field near the phase transition and will not attempt to compute the critical exponents of the theory. It is then
possible to justify the previously assumed dependence of the mass of the fermion-antifermion scalar field on the number of flavors using 
the trace anomaly as supplementary information. The effective potential of the theory is 
\begin{eqnarray}
V[\sigma] = \frac{1}{2}M^2_{\sigma}(N_{Tf})\, \sigma^2 + \frac{\lambda}{4}\sigma^4 \ ,
\end{eqnarray}
while the kinetic term is normalized canonically. Since we are in the spontaneously 
broken phase of the theory $M^2_{\sigma}<0$ while the physical squared mass is proportional to the absolute 
value of $M^2_{\sigma}$. The associated trace of the energy momentum tensor is
\begin{eqnarray}
\theta^{\mu}_{\mu} = -M^2_{\sigma}(N_{Tf}) \sigma^2 \ .
\end{eqnarray}
Now we recall that the expression for the trace anomaly of the underlying $SU(N)$ gauge theory is
\begin{eqnarray}
\theta^{\mu}_{\mu} = -\frac{\beta}{2g} G^{\mu\nu;a}G_{\mu\nu}^a \ , \quad {\rm with} \quad {a=1,\cdots, N^2-1} \ . 
\end{eqnarray}
If the theory develops a conformal fixed point, the beta function vanishes. In order to connect the previous equation to the phase transition we are interested in we consider the 
two loop beta function 
\begin{eqnarray}\beta&=&-{\beta_0}\frac{g^3}{16\pi^2} -{\beta_1}\frac{g^5}{(16\pi^2
)^2} \ ,\quad 
\beta_0 = \frac{11}{3}N - \frac{2}{3}N_{T f}\,(N+2) \ , \nonumber \\  
\beta_1 &=&\frac{34}{3}N^2 - N_{T f} \left(N+
2\right)\left[\frac{10}{3}N + \frac{2}{N}\left(N - 
1\right)\left(N + 2\right)\right] \ . \nonumber
\end{eqnarray}
Here we have provided the coefficients for a generic $SU(N)$ S-type theory with $N_{T f}$ flavors while 
the following result is independent of the specific representation to which the fermions belong. We 
rewrite the two loop beta function as follows:
\begin{eqnarray}
-\frac{\beta}{2g}=\frac{\beta_1}{32\pi^2}\,\alpha \left(\alpha - \alpha_{\ast}\right) \ , \quad {\rm with } \quad \frac{\alpha_{\ast}}{4\pi} = -\frac{\beta_0}{\beta_1} \ .
\end{eqnarray}
As we decrease $N_{Tf}$ relative to $N$ we have that $\alpha_{\ast}$ increases. 
To extract further information we impose the extra condition that the anomalous dimension $\gamma$ of the quark operator near the phase transition assumes the value one.
This last condition is 
consistent with the fact that one expects chiral 
symmetry to be broken for $\gamma>1$ \cite{{Cohen:1988sq},{Hill:2002ap}}. Define with $\alpha_c$ the special value of the coupling constant for which the previous condition on the anomalous dimension is satisfied. We have then:
 \begin{eqnarray}
-\frac{\beta[\gamma=1]}{2g_c}=\frac{\beta_1}{32\pi^2}\,\alpha_c \left(\alpha_c - \alpha_{\ast}\right) \propto N^c_{Tf} - N_{Tf}\ .
\end{eqnarray}
{}For $\alpha_{\ast}$ above $\alpha_{c}$ chiral symmetry breaks. So, in order to compare with the effective Lagrangian theory we are  
approaching $\alpha_c$ from large values of $\alpha_{\ast}$. However the previous expression is valid on both sides of the transition. 
If the two trace anomalies, i.e. the one in the effective theory and the one in the underlying theory, are describing the same physics then 
the flavor dependence is contained in the mass of the scalar field which reads near the phase transition:
\begin{eqnarray}
|M^2_{\sigma}(N_f)|\propto (N^c_{Tf} - N_{Tf}) \ .  
\end{eqnarray}
Interestingly this is exactly the mean field theory type relation, where $N_{Tf}$ is the scaling parameter of the theory. The previous way of computing the dependence of the mass of the fermion-antifermion field on the number of flavors near the fixed point  is probably too crude of an approximation, but 
it fits well with the mean field theory approach to a phase transition. Besides, never in 
our previous approach we needed perturbation theory since we can always work in the 't Hooft scheme in which the 
two loop beta function is exact. It is only when we try to provide a value 
for the critical number of flavors that approximations must be made in non-supersymmetric theories. 
We expect corrections to the scaling behavior reported above. Besides, non-perturbative computations in the underlying 
theory \cite{Appelquist:1991kn} indicate that the chiral condensate vanishes very fast near the transition as function of the
number of flavors. This behavior has been captured in \cite{Sannino:1999qe}. Here to provide the exponentially fast decrease of the chiral condensate one 
had to incorporate the axial anomaly of the underlying theory at the effective Lagrangian level. This then generated a nonanalytic potential for the 
effective low energy theory. In any case the flavor dependence of the Higgs mass provided in the present paper can be considered as a conservative estimate since it might drop even faster when approaching the fixed point.

Another possible criticism would be that near the conformal phase transition other states are expected to 
become light. Although this may affect the argument supporting universal 
behavior near the phase transition, as was already stressed by 
Chivukula \cite{Chivukula:1996kg}, it needs not to influence our result. 
This is so since the trace anomaly can be saturated by a single field and new fields can be added in the effective theory in a way that 
they do not contribute to the trace anomaly of the underlying theory. 
See, for example the extended Veneziano-Yankielowicz theory \cite{Veneziano:1982ah,{Merlatti:2004df}} which now contains gluinoballs and glueballs but the anomaly is saturated only by the 
gluinoball fields as in the original theory.

We observe that the mass of the scalar is reduced from its value at $N_{Tf}=1$ as $N_f$ increases. While a great deal
is known about the phase diagram of supersymmetric gauge theories
as function of the number of flavors and colors, much less is known
about the non-supersymmetric gauge theories. 
The perturbative estimate for $N_f^c$ in the S-type theories, in the case of the two-index
symmetric representation as a function of the number of colors was computed in \cite{Sannino:2004qp} and it is
$N_{Tf}^c={83N^3+66N^2-132N}/({20N^3+55N^2-60})$. 
{}For the two(three) technicolor theory it yields $N_f^c\simeq 2.1(2.5 )$. If we use
as $M_\sigma(N_{Tf}=1)=300 - 500$~GeV, we obtain
\begin{eqnarray}
m_{H}=M_\sigma(N_{Tf}=2)&\sim &90 - 150~{\rm GeV}  \ , \qquad {\rm S-two~technicolors} \\ ~&&\\ m_{H}=M_\sigma(N_{Tf}=2)&\sim&170 - 300~{\rm GeV}  \ , \qquad {\rm S-three~technicolors} . 
\end{eqnarray}
Note that even if we would have chosen $1$~TeV as normalization mass for one flavor the mass of the scalar near the phase transition would still be highly suppressed. Here $M_{\sigma}$ denotes the physical mass of the scalar meson.
Interestingly, not only are we able to resolve the ordinary hierarchy problem but we can also account for a 
dynamical light Higgs. 
\begin{table}[h]
\begin{tabular}{c|c|c|c|c|c}
   & QCD-like & WTC$(3,11)$ &WTC$(2,7)$ &S$(3,2)$ & S$(2,2)$ \\
  \hline \hline
  $m_H$(GeV)~$\approx$ & $1000$ & $400$ & $300$ & $170-300$ & $90-150$ \\
  \hline \hline
\end{tabular}
\caption{Higgs mass for QCD-like theories, walking technicolor with fermions in the fundamental representation WTC$(N,N_{TF})$, and for the S-type theories S$(N,N_{Tf})$.}  
\label{tableMh}
\end{table}

{}For comparison we evaluate the mass of the scalar field (i.e. Higgs field) in the case of ordinary walking technicolor with fermions in 
the fundamental representation of the gauge group. Here we use as normalization point the case of three flavors and colors which is the 
scaled up version of QCD. In these theories the critical number of flavors is $N^c_{T f}\approx 3.9 N$ which leads to the value of the Higgs near the transition of the order of $m_{H}\approx 290$~GeV for three colors and eleven flavors. In the case of the 
two color theory with seven 
flavors while still assuming a mass of the order of one TeV for the two flavor case one obtains $m_{H} \approx 370$~GeV. In the case of the fermions in the fundamental representation of the gauge group, due to the large value of the flavors needed to approach the conformal fixed point, our approximations are expected to be less reliable. In table \ref{tableMh} we summarize the expected masses for the Higgs for different types of technicolor theories.


\section{Conclusion}
\label{8}

We investigated theories of composite Higgs with technifermions in higher dimensional representations of the technicolor 
gauge group. More specifically, we studied theories in which the technifermions are in 
the two-index symmetric representation of the technicolor group. These theories have the important property that they are already quasi
 conformal with a very low number of technifermions. Somewhat surprisingly 
the lowest number of techniflavors needed to render the theory quasi conformal is {\it just} two, i.e. 
exactly one doublet of technifermions with respect to the weak interactions. This fact promotes these theories to ideal candidates of 
technicolor models of ``walking'' type.  
{}From the point of view of the weak interactions, the two technicolor theory with two techniflavors has a Witten anomaly which we cured 
by introducing a fourth family of heavy leptons. Among the interesting features of this model is 
that from the electroweak
 theory point of view the techniquarks resemble exactly one new family of quarks only rearranged in a higher 
representation of the $SU(2)$ technicolor theory. We have also studied the theory with three technicolors and 
two techniflavors in the two-index symmetric representation
of the gauge group which does not require a fourth family of leptons. 

{We have introduced theories with a critical number of flavors needed to enter the conformal window 
which is higher than in the one with fermions in the two-index symmetric representation but lower than in the 
traditional walking technicolor models. 
A simple class of these theories are split technicolor theories in which 
 we add only a (techni)gluino in the theory with still $N_{Tf}$ fermions in the fundamental representation of the 
gauge group. These theories have 
 a number of features in common with the ones used in 
 recent extensions of the standard model \cite{Arkani-Hamed:2004fb,{Giudice:2004tc},{Arkani-Hamed:2004yi}}. 
However, they were introduced here to address the hierarchy problem.}

We have then confronted the theories with fermions in the two index symmetric representation with the recent electroweak precision
measurements and have demonstrated that the theory with two technicolors is a natural candidate of electroweak symmetry breaking via new strong interactions. The theory with three technicolors and degenerate technifermions is less favored by the electroweak 
precision measurements. We have also briefly investigated the technibaryon spectrum in both theories which can lead to interesting distinctive signatures at LHC.

Amusingly, we also identified possible candidates for cold dark matter such as a new heavy neutrino. This 
particle is constrained to have a mass of the order of the neutral weak gauge boson and has an associated charged 
lepton with a mass roughly twice as large. It would be very interesting to carry out calculations for the distribution 
of dark matter for such a candidate along the lines of \cite{Green:2005fa}. One could thus gain further 
information on our model from cosmological observations \cite{Spergel:2003cb,Tegmark:2003ud}. We might also have 
possible candidates due to neutral technibaryons but in this case a more complete analysis is needed 
\cite{Chivukula:1989qb}.

A number of possible future directions can be outlined. It is relevant to construct the low-energy effective theories. 
Due to the presence of a light Higgs the underlying theory may naturally yield the higher-dimension operators added 
in effective Lagrangian descriptions of the electroweak symmetry breaking built in the past on more 
phenomenological grounds \cite{Barbieri:1999tm,{Bagger:1999te},{Kolda:2000wi},{Chivukula:2000px}} in order to be 
able to fit the electroweak precision measurements.
Thanks to a partial resemblance with supersymmetric Yang-Mills theory as well as the near quantum phase 
transition properties such as near parity doubling are expected in the technihadronic spectrum.  The underlying 
theories may then produce low-energy effective theories sharing some features with 
the more phenomenological models presented in \cite{Casalbuoni:1995qt}. 

{It is tempting to speculate that the cosmological constant problem \cite{Spergel:2003cb,{Tegmark:2003ud}} may be resolved by assuming a nearby 
quantum phase transition for the universe. In this scenario the cosmological constant is expected to be the function 
of a yet unknown scaling parameter, similar to the number of flavors for the Higgs mass, which turns out to be 
very close to its critical value.}

Lattice simulations of QCD are becoming more reliable in computing the coefficients of the low-energy effective Lagrangian for QCD 
as stressed in Gupta's review talk \cite{Gupta:2005vz}. At least two lattice collaborations, HPQCD-MILC-UKQCD \cite{Aubin:2004ck} and CP-PACS/JLQCD \cite{Ishikawa:2004xq} have 
reported important results by simulating $2+1$ flavors with dynamical fermions. It should then already be 
possible to determine the coefficients of the low-energy effective theories discussed above for 
the strongly interacting theories presented here. {The determination of the running of the coupling constant via lattice \cite{Luscher:1993gh} 
is also a test of our predictions.}
These computations will not only help to make further quantitative predictions for LHC but will also help in 
understanding strong dynamics of which QCD is a single but time-honored and physical example.

While we have assumed a bottom-up approach and constructed explicit technicolor theories which are not ruled out by 
the current precision measurements, it is also very important, at this point, to consider extensions capable of 
addressing the mass generation problem in some detail. A natural step would be to consider extending recent 
work \cite{Appelquist:2002me,Appelquist:2003uu,{Appelquist:2003hn},{Appelquist:2004mn},{Appelquist:2004es},{Appelquist:2004ai},{Martin:2004ec}}
concentrated on extended technicolor theories and their phenomenological impact for walking technicolor with technifermions 
in the fundamental representation of the gauge group to the theories presented here which are shown not to be at 
odds with precision data. Strong and weak CP violation might also be explained in technicolor-like extensions of 
the standard model \cite{Lane:2001rv,{Hsu:2004mf}}. Also the work in \cite{Giudice:1991sz,Doff:2004mm,{Doff:2003qc},{Doff:2003by},{Doff:2002kq}} can be extended to the present type of technicolor theories. We also expect our results to be useful when constructing technicolor models \cite{Haba:2005jq}.  

The unification of gauge couplings has not been addressed here but will be investigated in the near future \footnote{{}For example it
would be interesting, among other things, to explore the ideas presented in \cite{Grunberg:1987nc,{Grunberg:1987vy}} and 
related to the work in \cite{Maiani:1977cg}.}. 
{}For this purpose, it is then interesting to observe that the two technicolor theory can 
be rewritten as an $SO(3)$ theory with fermions in the fundamental representation of the gauge group.

The finite temperature electroweak phase transition with the present theories is also an interesting avenue to 
explore. The outcome is relevant for the baryogenesis problem since the baryon asymmetry cannot be explained 
within the standard model. Not only is there not enough CP violation \cite{Gavela:1994dt}, but the phase transition is a 
smooth crossover for Higgs masses larger than $80$~GeV \cite{Kajantie:1996mn}. However, recently it has been 
argued that the presence of dimension six operators in the Higgs Lagrangian can help in driving the transition to 
being of first order \cite{Bodeker:2004ws}. We point out that the presence of 
higher-order operators of the type suggested in the literature are naturally expected to emerge when breaking the 
electroweak theory using the technicolor theories presented here. The cut-off in the Higgs Lagrangian seen as a low energy effective theory, is naturally identified with the mass of the excited hadronic states which have been integrated out at low energy. The resulting 
effective theory would emerge in a fashion very similar to the toy model presented in \cite{Grojean:2004xa}.

Electroweak baryogenesis will also be affected by the presence of a new lepton family. Interestingly the new family is predicted to be visible at LHC and the underlying theories could be a natural starting point for models of electroweak baryogenesis proposed recently \cite{Abada:2004wn,Hall:2005aq}.

In short we have provided technicolor theories which are not ruled out by electroweak precision measurements, 
naturally yield 
a very light Higgs while predicting the existence of a new fourth lepton family with masses of the order of the 
electroweak 
scale. Since they are near conformal with just two techniflavors the dangerous flavor changing neutral currents are suppressed while having only 
few or no unwanted pseudogoldstone bosons. The Higgs is light due to a nearby quantum phase transition. We have developed a way to estimate its mass for different technicolor theories and shown that some of the theories presented here have a composite Higgs with a mass less than or equal to $150$~GeV. Possible cosmology oriented applications have been suggested. 

\acknowledgments
We would like to thank J.H.~Bijnens, S.~Chivukula, P.H.~Damgaard, L.~Del Debbio, 
P.~Di Vecchia, C.~Froggat, P.~Hansen, S.~Hofmann, D.K.~Hong, S.D.~Hsu, A.D.~Jackson, C. Jarlskog, P. Merlatti, H.B.~Nielsen, P.~Olesen, T.A.~Ryttov, J.~Schechter, M.~Shifman, R.~Shrock, 
B.~Svetitsky, D.J. Schwarz, and L.C.R.~Wijewardhana for helpful discussions, comments or careful reading of the manuscript. %

\noindent
The work of F.S. is supported by the Marie Curie Excellence Grant as team leader under contract MEXT-CT-2004-013510 and as Skou fellow of the Danish Research Agency. K.T. thanks the Academy of Finland for support under grant No. 206024.

\appendix
\section{Summary of the
 Formulae we used to compute the Oblique Parameters}

The oblique parameters $S$, $T$, and $U$ are defined as in \cite{He:2001tp}:
 \be S&=&-16\pi\frac{\Pi_{3Y}(m_Z^2)-\Pi_{3Y}(0)}{m_Z^2} \,, \nn T&=&4\pi\frac{\Pi_{11}(0)-\Pi_{33}(0)}{s_W^2c_W^2m_Z^2} \,, \nn U&=&16\pi\frac{[\Pi_{11}(m_Z^2)-\Pi_{11}(0)]
              -[\Pi_{33}(m_Z^2)-\Pi_{33}(0)]}{m_Z^2}
\ee
where the weak-mixing angle $\theta_W$ is defined at the scale $\mu =m_Z$.
$\Pi_{11}$ and  $\Pi_{33}$ are the vacuum polarizations of isospin currents, and $\Pi_{3Y}$ the vacuum polarization of one isospin and one hypercharge current.

In the case of Dirac-type masses for the fermions the one-loop fermionic contributions to the oblique parameters read
\cite{He:2001tp}:
\begin{eqnarray}
S_f&= & \frac{\sharp}{6\pi}
\left\{
2(4Y+3)x_1+2(-4Y+3)x_2-2Y\ln\frac{x_1}{x_2}
\right.\nonumber \\[1mm]
&&\left. +\left[\left(\frac{3}{2}+2Y\right)x_1+Y\right] G(x_1)  +\left[\left(\frac{3}{2}-2Y\right)x_2-Y\right] G(x_2) \right\}\,, \label{eq:Sfermion}\\[3mm]
T_f&=&\frac{\sharp}{8\pi{s}_W^2c_W^2}F(x_1,x_2) \,, \label{eq:Tfermion}\\[3mm] U_f&=&-\frac{\sharp}{2\pi}\left\{\frac{x_1+x_2}{2}-\frac{(x_1-x_2)^2}{3}
+\left[\frac{(x_1-x_2)^3}{6}-\frac{1}{2}\,\frac{x_1^2+x_2^2}{x_1-x_2}
\right]
\ln\frac{x_1}{x_2}\right. \nonumber\\[1mm] &&+\left.\frac{x_1-1}{6}f(x_1,x_1)+\frac{x_2-1}{6}f(x_2,x_2)+
\left[\frac{1}{3}-\frac{x_1+x_2}{6}-
\frac{(x_1-x_2)^2}{6}\right]f(x_1,x_2)\right\} , \label{eq:Ufermion} \end{eqnarray} where  $x_i=(M_i/m_Z)^2$ 
with $i=1,2$ and the factor $\sharp = N$  for techniquarks in the fundamental representation of the $SU(N)$ gauge group, $\sharp = N(N+1)/2$ for techniquarks in the 
two-index symmetric representation of 
the gauge group, and $\sharp=1$ for one doublet, for example of leptons.
The functions $G(x)$, $F(x_1,x_2)$, and $f(x_1,x_2)$ are defined via:

\begin{eqnarray}
F(x_1,x_2)&=&\frac{x_1+x_2}{2}-\frac{x_1x_2}{x_1-x_2}\ln\frac{x_1}{x_2}
\,,
\label{eq:Ffun}
\\[2mm]
G(x)&=&-4\sqrt{4x-1}\,\arctan\frac{1}{\sqrt{4x-1}}
\,,
\label{eq:Gfun}
\\[2mm]
f(x_1,x_2)&=&\left\{
\begin{array}{ll}
-2\sqrt{\Delta}\left[\arctan\frac{x_1-x_2+1}{\sqrt{\Delta}}
-\arctan\frac{x_1-x_2-1}{\sqrt{\Delta}}\right]\,,
&~~(\Delta>0)\,,  \\[1.5mm]
0\,,
&~~(\Delta=0)\,,\\[1.5mm]
\sqrt{-\Delta}\ln\frac{x_1+x_2-1+\sqrt{-\Delta}}{x_1+x_2-1-\sqrt{-\Delta}}\,,
&~~(\Delta<0)\,,
\end{array}
\right.
\label{eq:ffun}\\[3mm]
\Delta&=&2(x_1+x_2)-(x_1-x_2)^2-1  \, .
\end{eqnarray}
{ We note that recently it has been shown that for fermion masses near half of the neutral weak vector boson mass one needs corrections \cite{Novikov:2002tk}.}

\end{document}